\documentclass[journal]{IEEEtran}
\usepackage{amsfonts}
\usepackage{amssymb}
\usepackage[cmex10]{amsmath}
\usepackage{graphicx}
\usepackage{subfigure}
\usepackage{cite}
\usepackage{siunitx}
\usepackage{tabularx,booktabs}
\usepackage{amsthm}
\usepackage{booktabs}
\usepackage[ruled,lined]{algorithm2e}
\usepackage{multirow}
\usepackage{setspace}
\usepackage{fancyhdr}
\usepackage{algorithmic}
\usepackage{enumerate}
\usepackage{color}
\usepackage{epstopdf}
\usepackage{multicol}
\usepackage{setspace}
\begin{document}
\newcommand{\be}{\begin{equation}}
\newcommand{\ee}{\end{equation}}
\newcommand{\br}{{\mbox{\boldmath{$r$}}}}
\newcommand{\bp}{{\mbox{\boldmath{$p$}}}}
\newcommand{\bpi}{\mbox{\boldmath{ $\pi $}}}
\newcommand{\bn}{{\mbox{\boldmath{$n$}}}}
\newcommand{\balfa}{{\mbox{\boldmath{$\alpha$}}}}
\newcommand{\ba}{\mbox{\boldmath{$a $}}}
\newcommand{\bta}{\mbox{\boldmath{$\beta $}}}
\newcommand{\bg}{\mbox{\boldmath{$g $}}}
\newcommand{\bPsi}{\mbox{\boldmath{$\Psi $}}}
\newcommand{\bsigma}{\mbox{\boldmath{ $\Sigma $}}}
\newcommand{\bGamma}{{\bf \Gamma }}
\newcommand{\bA}{{\bf A }}
\newcommand{\bP}{{\bf P }}
\newcommand{\bX}{{\bf X }}
\newcommand{\bI}{{\bf I }}
\newcommand{\bR}{{\bf R }}
\newcommand{\bZ}{{\bf Z }}
\newcommand{\bz}{{\bf z }}
\newcommand{\bx}{{\mathbf{x}}}
\newcommand{\bM}{{\bf M}}
\newcommand{\bU}{{\bf U}}
\newcommand{\bD}{{\bf D}}
\newcommand{\bJ}{{\bf J}}
\newcommand{\bH}{{\bf H}}
\newcommand{\bK}{{\bf K}}
\newcommand{\bm}{{\bf m}}
\newcommand{\bN}{{\bf N}}
\newcommand{\bC}{{\bf C}}
\newcommand{\bL}{{\bf L}}
\newcommand{\bF}{{\bf F}}
\newcommand{\bv}{{\bf v}}
\newcommand{\bSigma}{{\bf \Sigma}}
\newcommand{\bS}{{\bf S}}
\newcommand{\bs}{{\bf s}}
\newcommand{\bO}{{\bf O}}
\newcommand{\bQ}{{\bf Q}}
\newcommand{\btr}{{\mbox{\boldmath{$tr$}}}}
\newcommand{\bNSCM}{{\bf NSCM}}
\newcommand{\barg}{{\bf arg}}
\newcommand{\bmax}{{\bf max}}
\newcommand{\test}{\mbox{$
\begin{array}{c}
\stackrel{ \stackrel{\textstyle H_1}{\textstyle >} } { \stackrel{\textstyle <}{\textstyle H_0} }
\end{array}
$}}
\newcommand{\tabincell}[2]{\begin{tabular}{@{}#1@{}}#2\end{tabular}}
\newtheorem{Def}{Definition}
\newtheorem{Pro}{Proposition}
\newtheorem{Lem}{Lemma}
\newtheorem{Exa}{Example}
\newtheorem{Rem}{Remark}
\newtheorem{Cor}{Corollary}
\renewcommand{\labelitemi}{$\bullet$}

\title{Heterogeneous Multi-sensor Fusion with Random Finite Set Multi-object Densities}
\author{Wei~Yi,~\IEEEmembership{Member,~IEEE},
        and Lei~Chai,~\IEEEmembership{Student Member,~IEEE}
\thanks{This work was supported in part by the National Natural Science Foundation of China under Grants 61771110, in part by the Fundamental Research Funds of Central Universities ZYGX2020ZB029, in part by Chang Jiang Scholars Program, and in part by the 111 Project B17008. }
\thanks{Wei Yi, and Lei Chai are with the School of Information and Communication Engineering, University of Electronic Science and Technology of China. (e-mail: kussoyi@gmail.com; yutian0113cl@gmail.com.)
}}

\markboth{IEEE Transactions on Signal Processing}%
 {Shell\MakeLowercase{\textit{et al.}}: Bare Demo of IEEEtran.cls for Journals}

\maketitle
\begin{abstract}
This paper addresses the density based multi-sensor cooperative fusion using random finite set (RFS) type multi-object densities (MODs). Existing fusion methods use scalar weights to characterize the relative information confidence among the local MODs, and in this way the portion of contribution of each local MOD to the fused global MOD can be tuned via adjusting these weights. Our analysis shows that the fusion mechanism of using a scalar coefficient can be oversimplified for practical scenarios, as the information confidence of an MOD is complex and usually space-varying due to the imperfection of sensor ability and the various impacts from surveillance environment. Consequently, severe fusion performance degradation can be observed when these scalar weights fail to reflect the actual situation. We make two contributions towards addressing this problem. Firstly, we propose a novel heterogeneous fusion method to perform the information averaging among local RFS MODs. By factorizing each local MODs into a number of smaller size sub-MODs, it can transform the original complicated fusion problem into a much easier parallelizable multi-cluster fusion problem. Secondly, as the proposed fusion strategy is a general procedure without any particular model assumptions, we further derive the detailed heterogeneous fusion equations, with centralized network architecture, for both the probability hypothesis density (PHD) filter and the multi-Bernoulli (MB) filter. The Gaussian mixture implementations of the proposed fusion algorithms are also presented. Various numerical experiments are designed to demonstrate the efficacy of the proposed fusion methods.

\end{abstract}
\IEEEpeerreviewmaketitle

\begin{table}[h!]
	\renewcommand{\arraystretch}{1.2}
	\begin{center}

		\begin{tabular*}{0.48\textwidth}{@{\extracolsep{\fill}}l l}
			\multicolumn{2}{c}{\textsc{List of Acronyms}}
\vspace{2mm}\\
			AEUF & Accuracy Estimation Uncertainty Function \\
			CEUF & Cardinality Estimation Uncertainty Function \\
				CPHD & Cardinalized Probability Hypothesis Density                       \\
			EUF &  Estimation Uncertainty Function                     \\
			FoV & Field of View                     \\
			GLMB & Generalized Labeled Multi-Bernoulli \\
			GM & Gaussian Mixture \\
			H-MMB & Heterogeneous Multi-sensor Multi-Bernoulli\\
			H-MPHD & Heterogeneous Multi-sensor PHD \\
			HPD & Highest Posterior Density\\
			ICC & Information Confidence Coefficient \\
			KLD & Kullback-Leibler Divergence \\
			MB & Multi-Bernoulli \\
			MOD & Multi-object Density \\
			MPP & Multi-object Poisson Process \\
			OSPA & Optimal Sub-pattern Assignment  \\
			PHD & Probability Hypothesis Density \\
			RFS & Random Finite Set \\
			WAA & Weighted Arithmetical Average \\
			WGA & Weighted Geometrical Average\\
		\end{tabular*}
		\label{tabone}
		\normalsize
	\end{center}
\end{table}

\section{Introduction}
\label{sec:intro}

\IEEEPARstart{M}{ulti-object} estimation involves the simultaneous estimation of a time-varying number of objects and their states from a set of noisy sensor observations. This problem arises in many fields, particularly, the multi-object tracking in radar, sonar and camera surveillance. The \emph{multi-object density} (MOD) is of fundamental importance in multi-object estimation as it statistically captures the information on the number of objects and the states of individual objects.

By modeling multi-object states and sensor observations as \emph{random finite sets} (RFSs), the Bayesian inference of the RFS type MODs provides a unified mathematical framework for multi-object estimation~\cite{Mahler2007Book}. The resultant multi-object Bayes filter can be viewed as a systematic generalization of the Kalman type single-target Bayes filter. However, the full RFS multi-object filter can be computationally intractable for most practical scenarios due to its high dimensionality, and thus efficient approximations are needed. Generally, there are two kinds of approximations, namely, the moment-type and parameter-type approximations.

The probability hypothesis density (PHD) filter~\cite{MahlerPHD,Vo2006} and the cardinalized PHD (CPHD) filter~ \cite{MahlerCPHD,vo2007analytic} belong to the moment-type approximation, and they propagate, respectively, the MODs of {{multi-object Poisson processes}} (MPPs) and {independent identically distributed (i.i.d.) cluster processes} forward in the Bayesian recursions. The majority of the rest filters belong to the second kind, which propagates forward the parameters that characterizing the corresponding MODs in the Bayesian recursions. Examples include the multi-Bernoulli (MB) filter~\cite{Mahler2007Book,Vo2009,Vo2010}, the labeled MB filter~\cite{Reuter2014}, and the generalized labeled multi-Bernoulli (GLMB) filter~\cite{Vo2013,Vo2014}. Due to the concise and rigorous mathematical representation of the RFS approaches, these filters have attracted increasing interests in both academic and industrial fields.

The recent rapid advances in wireless communication and sensor technology have accelerated the transformation of modern sensing systems. One important trend is networking. Indeed, combining multiple sensors, possibly with different types and abilities, into a collaborative network can yield enhanced multi-object estimation performance compared to the isolated individuals. Since MOD statistically captures the information on the number of objects and their states, the MOD based multi-sensor cooperative fusion is an effective solution, and has been widely studied and used in many defense and civilian areas. Its essential question is how to integrate multiple MODs from local sensors into a fused global MOD in a proper and efficient manner. In this regard, the fusion methods can have decisive impacts to the final estimation performance.

By now, the two commonly adopted methods for fusing MODs are the generalized covariance intersection~\cite{MahlerGCI2000,Ueny2019_TAES}, or exponential mixture densities~\cite{EMD-Julier2006,ClarkGCI2010} and the {arithmetic mixture densities}~\cite{Bailey2012,Yu2016_ICASSP,LTC_2017_Fusion}, or minimum information loss~\cite{gao2020multiobject}. In both methods, a scalar weight is assigned to each local MOD before fusing in order to characterize its relative \emph{information confidence}\footnote{Here we use a notion of {information confidence} to account for how reliable a certain amount of information contained in an MOD is, and it is determined by the specific sensor and environmental conditions under which the information is obtained. We will show how the information confidence of an MOD can be quantitatively evaluated later in Section V-B.} compared with other local MODs. The fused MODs by the two methods are, respectively, the \emph{weighted geometrical average} (WGA) and the \emph{weighted arithmetic average} (WAA) of the local MODs. In the rest of the paper, we refer to these two methods as WGA and WAA for terminology simplicity. It has been shown that both methods are immune to the double counting of common information due to the unknown correlations of measurements~\cite{EMD-Julier2006,Bailey2012,LTC2019}, and can provide better situation awareness compared to each local MOD. Nevertheless, it should be also noted that both the WGA and WAA fusion rules are not optimal since they do not attempt to calculate and cancel the common information during the fusion~\cite{EMD-Julier2006,li2020best}.

In the recent years, tractable implementations of WGA and WAA have been derived for various RFS MODs. In~\cite{ClarkGCI2010}, the analytical fusion equations of WGA were derived for the MPP and i.i.d. cluster process MODs. Later on, WGA has been successfully applied to various RFS filters, including the PHD~\cite{Uney2013}, CPHD~\cite{Battistelli2013}, MB~\cite{Yi2020TSP,Wang2017} filters and most recently labeled RFS filters~\cite{Suqi2018,Suqi2019}. As for WAA, a multi-sensor PHD filter is derived analytically in~\cite{Reza2017}.  The applications of WAA to other RFS filters are still under investigation and the recent studies can be found in~\cite{li2020arithmetic,gao2020multiobject,Kai2019,Gao_2020_TSP}. It has been shown that
the fused MODs by WGA and WAA minimize, respectively, the information gain and information loss from the local MODs~\cite{Kullback1951,Battistelli2013,Kai2019,Giorgio_KLA,gao2020multiobject}. 

As mentioned before, a scalar weight is assigned to each local MOD in order to characterize its relative information confidence in both WGA and WAA. In this way, we can tune, through these scalar weights, the portion of contribution of each local MOD to the fused global MOD. However, this fusion mechanism of using a scalar coefficient can be oversimplified for many practical scenarios, as the information confidence of an MOD is usually space-varying due to the limited sensor ability and various impacts from sensing environment. For example, a radar sensor can have limited detection range due to the radiation attenuation, and the sensing horizon of a sensor can be blocked by sight obstructions. In this case, the information confidence of an MOD cannot be sufficiently characterized using only a scalar coefficient, and severe fusion performance degradation can be observed due to the weighting bias~\cite{FoV3}. This problem is especially evident in the fusion among sensors having different fields of view (FoVs), since the relative information confidence is different in different regions due to the non-overlapping FoVs~\cite{Yi2020TSPDFoVs}. Recently, the different FoV issue has been intensively studied, and effective countermeasures have been proposed for both unlabel MODs~\cite{FoV2,FoV4,FoV5,Da_2020_TSIPN,Yi2020TSPDFoVs} and labelled RFS MODs~\cite{FoV6,FoV8,Gostar_2021_TSP}. However, these methods do not change the fundamental problem of the fusion mechanism.

In addition to the density based fusion methods, the measurement based fusion methods have also been extensively studied recently. In the later case, by assuming the measurements independence among local sensors, Bayes fusion methods using raw measurements can be derived to obtain the optimal estimation performance. Typical examples include the measurement based multi-sensor PHD filter~\cite{Mahler2010approximate}, CPHD filter~\cite{Nannuru2016multisensor}, MB filter~\cite{Saucan2017multisensor} and most recently GLMB filter~\cite{Vo2019multisensor}.

In this paper, we propose a novel fusion strategy to address the problem of information averaging among multiple RFS MODs with space-varying information confidence. {As an early attempt to address this problem, a heuristic remedy was proposed in~\cite{FoV3}, but neither general methodology nor analytical fusion equations were provided. Here we propose a novel fusion strategy along with its detailed implementations.} The main contributions can be summarized as follows:

\begin{enumerate}
  \item \emph{Heterogeneous fusion method for RFS MODs}: {Since the relative information confidence among local MODs is complex and space-varying, we propose to factorize each MOD into a number of smaller size sub-MODs by exploiting independence property of the RFS processes and applying the convolution formula~\cite{Mahler2007Book}. The purpose of the MOD factorization is to ensure that the information confidence of these sub-MODs can be sufficiently characterized using scalar fusion weights.} Then, by properly clustering sub-MODs into groups, the original complicated fusion problem is transformed into a parallelizable multi-cluster fusion problem, wherein each inner-cluster fusion can be executed, simply and independently, based on the tailor-made sub-MOD level weights. Lastly, the global fused MOD is obtained by combining the fused sub-MODs of all groups. {We refer to this strategy as \emph{heterogeneous fusion} since the information averaging involves multiple sets of the tailor-made fusion weights, each of which reflects the relative information confidence of the sub-MODs within a specific cluster and can be different for different clusters.}
  \item \emph{Analytical implementations using PHD and MB filters}: The proposed heterogeneous fusion strategy is a general procedure without any particular assumptions on the types of MODs. Therefore, in order to examine the efficacy of the heterogeneous fusion strategy, we further derive the detailed heterogeneous fusion equations for both MPP type and MB type MODs. The resultant filters are referred to as the heterogeneous multi-sensor PHD (H-MPHD) filter and the heterogeneous multi-sensor MB (H-MMB) filter, respectively. Both filters adopt centralized fusion architecture. Specifically, the local MODs of all sensor nodes are sent to a fusion center to compute a fused global MOD using the heterogeneous fusion equations, and there is no information feedback to each sensor. The rationale of the devised filters is analysed, and their relationship with the standard multi-sensor PHD and MB filters is also discussed. Lastly, the analytical expressions of both filters are given in the context of the Gaussian mixture (GM) implementation.
\end{enumerate}

The remainder of the paper is organized as follows. Section II introduces the notations and the background material. The motivation analysis is given in Section III. Section IV devotes to the development of the heterogeneous fusion method. The proposed H-MPHD and H-MMB filters are derived in Sections V and VI, respectively. Simulation results, demonstrating the efficacy of the proposed algorithms, are presented in Section VI. Finally, Section VII concludes the paper.

\section{Notations and Background}

We consider tracking a time-varying and unknown number of objects using a sensor network ${\mathcal{N}}$. The objects are assumed to move independently in a surveillance area. At each time step $k$, the multi-object state is represented as an RFS $X^k = \{ x^k_1,...,x^k_{n^k}\}$, where the variable $n^k$ denotes the number of objects and $x^k_j \in {\mathbb{X}}$ is the $j$-th single object state with ${\mathbb{X}}$ the single-object state space. The sensor network ${\mathcal{N}}$ consists of a number of spatially distributed sensor nodes, each of which has sensing and processing capabilities\footnote{{{In this paper, we assume that the transmission of the sensor network is reliable, i.e., message loss or corruption are not considered. Also, we assume all sensor nodes in the network have been correctly aligned in a common coordinate system either using global localization technologies or through self-localization methods such as~\cite{UMCcooperative2016,GBCWdistributed2020,Sharma2019_TSP}.}}}. At time $k$, by sensing the surveillance area, each sensor $i \in {\mathcal{N}}$, collects a set of measurements $Z^k_i = \{ z^k_{1,i},...,z^k_{o_i^k,i} \}$, where $o_i^k$ is the number of measurements and $z^k_{j,i} \in {\mathbb{Z}}$ with ${\mathbb{Z}}$ the measurement space.

The objective is to jointly estimate the number of objects and their states $X^k$ by combining the information of the whole sensor network. Let $\pi^k_{i}(X^k)$ denote the posterior MOD of sensor $i$ computed by the corresponding local filter on the basis of the collected measurements $Z^k_i$. Then, in the context of the density based multi-sensor multi-object estimation, the objective is achieved by obtaining a properly fused global MOD $\bar{\pi}^k(X^k)$ using $\{ \pi^k_{i}(X^k)\}_{i \in {\mathcal{N}}}$. In the following, the time index $k$ is dropped for the sake of
readability.


\subsection{RFS based Multi-object Densities}

From a Bayesian perspective, all the information on $X$ is contained in the corresponding posterior MOD $\pi$. In this paper, we focus our attention on the MPP and MB MODs as they are two fundamental and commonly used RFS MODs~\cite{Mahler2007Book}.

The PHD filter recursively propagates an MPP MOD forward. If $X$ is modelled as an MPP RFS, then the elements of $X$ are all independently identically distributed according to a same location density ${f\left( x \right)}$, and its cardinality distribution ${\text{Pr}(|X|=n)}$ is a Poisson distribution with parameter $\lambda$. The MOD of an MPP RFS is denoted as $\pi^{\text{\texttt{P}}} $ and given by~\cite{Mahler2007Book},
\begin{equation}\label{Poi distribution}
\pi^{\text{\texttt{P}}} \left( X \right) = {e^{ - \lambda }}\prod\limits_{x \in X}{\lambda}\cdot {f\left( x \right)}.
\end{equation}
Note that an MPP MOD is completely characterized by its PHD or intensity function $\nu^{\text{\texttt{P}}}(x)=\lambda{f\left( x \right)}$, which is defined on ${\mathbb{X}}$ and representing the first-order statistical moment of the MOD~\cite{Mahler2007Book,MahlerPHD,Vo2006}.

The MB filter recursively propagates an MB MOD forward. If $X$ is modelled as an MB RFS, then $X$ is defined as the union of a set of independent Bernoulli RFSs, i.e., $X=\bigcup_{b\in \mathbb{B}}X_b$, where $\mathbb{B}$ is the index set of the constituent Bernoulli components, and each Bernoulli RFS $X_b$ is characterized by its existence probability $r_{b}$ and state distribution $f_{b}(\cdot)$ conditioned on its existence~\cite{Mahler2007Book}. Let $\pi^{\text{\texttt{MB}}}$ denote the MOD of an MB RFS, then it is completely defined by the parameter set $\left\{\left(r_{b},f_{b}\right)\right\}_{b\in \mathbb{B}}$ and has the following form~\cite{Mahler2007Book},
\begin{align}\label{eq:MB_distribution}
\pi^{\text{\texttt{MB}}} (X) = \prod\limits_{b\in \mathbb{B}} \left( 1 - r_b\right)  \sum_{{{\scriptsize{\begin{array}{c}
{{l_1} \ne ... \ne {l_n} }   \\
{l_1},...,{l_n} \in \mathbb{B}
\end{array}}}}
}
\prod\limits_{j = 1}^n \frac{ r_{l_j} f_{l_j}(x_j) }{1-r_{l_j}} .
\end{align}
By definition~\cite[(16.66)]{Mahler2007Book}, the PHD of an MB is given by $\nu^{\text{\texttt{MB}}} (x)=\sum_{b\in \mathbb{B}} r_{b}f_{b}$. Note that the key spirit of MB RFS is its component-to-object pairing structure. Namely, each Bernoulli component is used to model the statistical uncertainties of a specific hypothesized object. This object-orientated structure also facilitates the extraction of the multi-object state estimate.

\subsection{MOD based Multi-sensor Fusion Rules}

Suppose that a set of MODs $\{ {\pi_i}\}_{i \in {\mathcal{N}}}$ are computed by the local filters in a network ${\mathcal{N}}$. The aim of multi-sensor fusion is to properly integrate $\{ {\pi_i}\}_{i \in {\mathcal{N}}}$ into a fused global MOD ${\bar{\pi}}$ which can provide better estimation performance. According to WGA~\cite{MahlerGCI2000,EMD-Julier2006,ClarkGCI2010}, the fused MOD is the geometric mean in the form of an exponential mixture of the local MODs,
\begin{equation}\label{eq:WGA}
\bar{\pi}(X) = \frac{\prod_{i\in {\mathcal{N}}} [\pi_i(X)]^{\omega_i}} {\int \prod_{i\in {\mathcal{N}}} [\pi_i(X)]^{\omega_i} \delta X},
\end{equation}	
where $\int \cdot\delta X$ denotes the set integral~\cite{Mahler2007Book}, and $\omega_i$ is a pre-set scalar fusion weight used to reflect the relative information confidence of the $i$-th MOD $\pi_i$ in comparison with the rest MODs and satisfies
\begin{equation}\label{weight}
{\sum}_{i\in\mathcal{N}}\omega_i=1, \hspace{3mm}\text{and} \hspace{3mm}\omega_i\geq 0.
\end{equation}
It has been shown that the fused MOD $\bar{\pi}$ of \eqref{eq:WGA}
actually minimizes the weighted sum of the Kullback-Leibler divergences (KLDs) from the local densities~ \cite{Kullback1951,Battistelli2013,Giorgio_KLA},
\begin{equation}\label{WGA PMDI}
\bar{\pi} = \mathop {\arg}\min\limits_\pi {\sum}_{i\in{\mathcal{N}}}{\omega_{i}}D_{\text{\texttt{KL}}}(\pi||\pi_i),
\end{equation}
where $D_{\text{\texttt{KL}}}(\cdot\|\cdot)$ denotes the KLD between two RFS MODs,
\begin{equation}\label{KLD}
D_{\text{\texttt{KL}}}(\pi_1||\pi_2) = \int \pi_1(X) {\log}{\frac{\pi_1(X)}{\pi_2(X)}} \delta X.
\end{equation}
The physical interpretation of (\ref{WGA PMDI}) is that the fused MOD $\bar{\pi}$ aims to minimize the discrimination information from the local ones as small as possible~\cite{Battistelli2013}.

Different from WGA, the fused MOD $\bar{\pi}$ via WAA is the arithmetic mean of the local MODs~\cite{Bailey2012,Reza2017,gao2020multiobject,Kai2019},
\begin{equation}\label{eq:WAA}
\bar{\pi}(X) = {\sum}_{i\in {\mathcal{N}}} \omega_i \pi_i(X).
\end{equation}
Interestingly, {it has been shown that WAA
in turn minimizes the weighted sum of the Cauchy Schwarz divergences (CSDs) with respect to the local MODs~\cite{Hoang2014,Reza2017}, and also minimizes the weighted sum of KLDs from the fused MOD to the local MODs~\cite{gao2020multiobject}}. Recently, it has been shown that the fused MODs of both WAA and WGA are Fr\'{e}chet means of local MODs, and they minimize their corresponding Fr\'{e}chet functions between the fused MOD and local ones as well~\cite{li2020arithmetic}.

\section{Motivation Analysis}\label{sec:motivation}

It can be seen from \eqref{eq:WGA} and \eqref{eq:WAA} that both fusion rules employ a set of scalar fusion weights $\{\omega_i\}_{i\in{\mathcal{N}}}$, each of which reflects the relative information confidence of the corresponding MOD compared with the rest ones. In other words, these fusion rules are theoretically feasible if and only if the relative information confidence of all local MODs $\{\pi_i\}_{i\in{\mathcal{N}}}$ is homogeneous on the state space. Here homogeneous means that the information confidence of an MOD remain unchanged for the entire state space, as only in that case their confidence can be sufficiently characterized by scalar weights. For this reason, we refer to \eqref{eq:WGA} and \eqref{eq:WAA} as \emph{homogeneous fusion strategy} in this paper.

However, the homogeneous assumption of the relative information confidence is generally too idealistic for many practical scenarios, and, consequently, severe fusion performance degradation is inevitable. To explain this more clearly, three commonly encountered scenarios are given as follows.

\emph{-- Example 1}: Radar-type sensors usually generate measurements in polar coordinates, i.e., the range-azimuth positions of objects with respect to the radar location. Due to its fixed angular sensing ability, the location estimation uncertainty of an object (in Cartesian coordinates) increases as the radar-object distance increases. Also, the detection uncertainty of an object raises as the object moves away from radar because of the radiation range attenuation. Clearly, the information confidence of a radar-type sensor based MOD is not homogeneous but space-varying along with the radar sensing ability.

\emph{-- Example 2}: In practical cases, sensors usually have limited FoVs, e.g., camera often have fixed square frames, and radars commonly have fan-shaped FoVs due to the radiation limitations. In~\cite{Yi2020TSPDFoVs}, theoretical analysis was given to show that direct homogeneous fusion among sensors with different FoVs is unsuitable and can result severe performance degradation. Recently a lot of effort has been devoted to devise the countermeasures so as to deal with the different FoV issue~\cite{FoV2,FoV4,FoV5,FoV8}. However, as is indicated in~\cite{Yi2020TSPDFoVs}, {the essence of the different FoV issue is that the relative information confidence among MODs is different in different regions due to the non-overlapping FoVs.} Thus it is unsuitable to average the MODs of sensors with different FoVs naively weighted by their inside FoV information confidence.

\emph{-- Example 3}: The environmental factors also have dramatic impacts on the information confidence of an MOD. For instance, the sensing horizon of a sensor equipped on mobile robots are time-varying and can be blocked by random and unknown obstructions. In this case, the information confidence of the MOD on different regions can change drastically, and consequently, is far from being accurate to be characterized using only a scalar fusion weight.

The above three examples indicate that the homogeneous fusion strategy has obvious limitations during the information averaging among RFS MODs. The main reason is that a sensor usually has different information confidence on different regions of its MOD, not to mention the more complex relative information confidence among the MODs of different sensors. Thus, it is oversimplified to homogeneously average MODs based on only a set of fixed scalar fusion weights.

\section{Heterogeneous Fusion Strategy}\label{sec:HFS}
In this section, we propose a novel fusion strategy to address the information averaging among MODs.
In the following, we first present the general framework of the fusion strategy without any particular assumptions on the types of MODs. In Sections \ref{sec:PHDimplem} and \ref{sec:MBimplem}, we will show how this strategy can be fulfilled with the MPP and MB MODs, respectively.

\subsection{General Methodology}\label{sec:generalm}

The ultimate goal of the proposed strategy is to achieve a reasonable information averaging among MODs by properly taking into account the complex and space-varying relative information confidence embedded in them. To this end, we propose a new fusion method and contains three major steps. Firstly, by exploiting the density structure, the MODs of local sensors are factorized into a number of smaller size sub-MODs, each of which concentrates only on a relatively small region of the state space. Then, the factorized sub-MODs of different sensors are matched into groups so that the information aggregation within each group is suitable. Also, tailor-made fusion weights are assigned to each group to properly reflect the relative information confidence among the sub-MODs in this group. Lastly, the fused global MOD defined on the entire state space is obtained by combining the fused sub-MODs of all groups.

{It can be seen that the key spirit of the proposed fusion method is a \emph{{local condition tailored}} tactic. Namely, by factorizing and clustering the local MODs into multiple group of sub-MODs, the information averaging can be performed more carefully and properly in group level based on multiple sets of tailor-made fusion weights, each of which can describe the local conditions of the relative information confidence within each sub-MODs cluster more accurately.} In this regard, we refer to the proposed strategy as a \emph{heterogeneous fusion} to emphasize its \emph{local condition tailored} way of information averaging.

\vspace{3mm}
\underline{\noindent {1) Decomposition of the local MODs }}
\vspace{3mm}

Consider that a set of MODs $\{\pi_i(X)\}_{i\in\mathcal{N}}$ are computed by the local filters on the basis of the measurements obtained at each sensor node. Each density $\pi_i(X)$ is then factorized into a number $M_i$ of sub-MODs $\{\pi_{i,m}(X_m)\}_{m=1}^{M_i}$ such that the following two conditions are satisfied:
\begin{itemize}
  \item C.1: {The factorization of $\pi_i(X)$ into $\{\pi_{i,m}(X_m)\}_{m=1}^{M_i}$ is based on the statistically independence between the random finite subsets $X_1,\ldots,X_{M_i}$ with $X = \cup_{m=1}^{M_i} X_m$.}
  \item C.2: {The information confidence within each sub-MOD $\pi_{i,m}$ needs to remain unchanged, namely, homogeneous.}
      \end{itemize}
When C.1 holds, the approximation error induced by MOD factorization is negligible, since the original MOD $\pi_i(X)$ can be reconstructed exactly using $\{\pi_{i,m}\}_{m=1}^{M_i}$ by applying the convolution formula~\cite[(11.252)]{Mahler2007Book},
\begin{equation}\label{eq:convo_decomp}
\pi_i(X) = \sum\limits_{ X_1 \biguplus \ldots \biguplus X_{M_i} = X } {\pi}_{i,1}(X_1)\cdots{\pi}_{i,{M_i}}(X_{M_i}),
\end{equation}
where the summation is taken over all mutually disjoint subsets $X_1,\ldots,X_{M_i}$ of $X$ such that $X = \cup_{m=1}^{M_i} X_m$.
To achieve C.1, the decomposition strategy needs to properly exploit the structure of the subject MOD, and can be different for different types of MODs. In Sections \ref{sec:PHDimplem} and \ref{sec:MBimplem}, two decomposition methods will be introduced for the MPP and MB MODs, which are the typical representatives of the moment-type and parameter-type MODs, respectively.

{When C.2 is satisfied, the information confidence of $\pi_{i,m}$ remains unchanged and thus can be sufficiently characterized using a scalar variable. Therefore, here we define a scalar-type information confidence coefficient (ICC), $\phi_{i,m}(x)\in[0,+\infty)$, and use it to characterize the homogeneous information confidence of $\pi_{i,m}$.} Since an MOD is computed on the basis of the sensor measurements, its information confidence is largely decided by the quality of its associated measurements. Therefore, C.2 holds when each factorized sub-MOD $\pi_{i,m}$ concentrates on a relatively small region of the state space such that the associated measurements have similar quality. On the other hand, the quality of measurements further depends on the sensor ability and the input effects of environment as well. For example, the estimation accuracy of object state is affected by the sensor resolution and measurement errors. The cardinality estimation on the number of objects can be affected by the sensor range attenuation and environmental impacts such as clutter intensity and obstructions of sensor sight line \cite{Yi2020TSPDFoVs}. Thus, the evaluation of ICC $\phi_{i,m}$ should take into account both sensor characteristics and environmental factors. {The detailed approach for the evaluation of the ICC will be discussed in the subsequent Sections V-B and VI-B.}

\vspace{3mm}
\underline{\noindent {2) Sub-density level information aggregation}}
\vspace{3mm}

Let $\{(\pi_{i,m},\phi_{i,m})\}_{m \in \mathbb{M}_i}$, with indexes $\mathbb{M}_i=\{1,\ldots,M_i\}$, denote the factorized sub-MODs and the corresponding ICCs of the $i$-th local MOD $\pi_i$. We now attempt to cluster all these sub-MODs into groups such that the information averaging within each group is feasible from the perspective of information theoretical view. Before proceeding, a formal Definition of a clustering of sub-MODs is given as follows.
\begin{Def}\label{clustering}
A clustering $\mathcal{C}=\{\mathcal{C}_1,\cdots,\mathcal{C}_{G_{\mathcal{C}}}\}$ of sub-MODs with density indexes $\left\{\mathbb{M}_i, {i  \in \mathcal{N}}\right\}$ is defined as a set of clusters, where the $g$-th, $g=1,\ldots,G_{\mathcal{C}}$, cluster is denoted as
$\mathcal{C}_{g}=\left\{\mathbb{M}_{i,g}, {i  \in \mathcal{N}}\right\}$. The term $\mathbb{M}_{i,g}$ is the index subset of sub-MODs grouped to form the $g$-th clusters and satisfies:
\begin{itemize}
	\item $\mathbb{M}_i=\cup_{g=1}^{G_{\mathcal{C}}} \mathbb{M}_{i,g}$, \hspace{2mm}${i  \in \mathcal{N}}$,
  \vspace{2mm}
    \item ${\cup}_{i  \in \mathcal{N}}\mathbb{M}_{i,g}\neq\emptyset$, \hspace{2mm}$g\in[1,\ldots,G_{\mathcal{C}}]$,
  \vspace{2mm}
    \item $\mathbb{M}_{i,g}\cap\mathbb{M}_{i,g'}=\emptyset$,\hspace{2mm}$(g,g')\in[1,\ldots,G_{\mathcal{C}}]^2 ,\hspace{1mm}g\neq g'. $
\end{itemize}
\end{Def}
{It can be seen from Definition $1$ that the index subsets $\mathbb{M}_{i,g}$ for all $g$ form a partition of $\mathbb{M}_{i}$ and are disjoint among each other.} Note that index subset $\mathbb{M}_{i,g}$ can be an emptyset. To ensure that the information fusion within each cluster is feasible, the clustering $\mathcal{C}$ of $\{(\pi_{i,m},\phi_{i,m})\}_{m \in \mathbb{M}_i}$ needs to satisfy the following two conditions:
\begin{itemize}
  \item C.3: The information contained in the sub-MODs of each cluster {needs to be coherent in the sense that these sub-MODs represent the statistical uncertainties of a same source, for example, a same object under tracking.}
  \item C.4: For any subset $\mathbb{M}_{i,g}$ contains more than one indexes, i.e., $|\mathbb{M}_{i,g}|>1$, {the information confidence of its indexed sub-MODs $\{\phi_{i,b}\}_{b\in \mathbb{M}_{i,g}}$ {are constant}}.
\end{itemize}
When C.3 holds, the information averaging within each cluster is feasible from the perspective of information consistency. To achieve C.3, the clustering can be performed by, for instance, judging the overlap extent of the highest posterior density (HPD) regions of sub-MODs~\cite{HPD2010,Wang2017,Yi2020TSPDFoVs}, or using density distances such as KLD~\cite{FoV4} or the divergence of generalized covariance intersection~\cite{Suqi2018,Suqi2019,Yi2020TSP} as clustering metrics. The necessity of C.4 is to ensure that the clustered sub-MODs from a same sensor have same \emph{local condition}, which is the prerequisite for the proposed \emph{local condition tailored} fusion tactic. Note that if both C.1 and C.2 are satisfied during the MOD factorization, each subset $\mathbb{M}_{i,g}$ usually only contains one density index or is an emptyset. This will be explained more clearly in Sections \ref{sec:PHDimplem} and \ref{sec:MBimplem}.

Up to now, the problem in front of us is how to perform an appropriate information averaging based on the clustered ${G_{\mathcal{C}}}$ groups of sub-MODs and their associated ICCs. Supported by C.3 and C.4, a natural and reasonable choice is to conduct homogeneous information fusion within each sub-MOD groups, but tailor-made fusion weights need to be designed beforehand. Let $\{\psi_{{i,g}}\}_{{i \in \mathcal{N}}}$ denote the ICCs of the sub-MODs in the $g$-th cluster, and we have $\psi_{{i,g}}=\phi_{i,b}$ for any $b\in\mathbb{M}_{i,g}$ due to C.4, and $\psi_{{i,g}}=0$ when $\mathbb{M}_{i,g}=\emptyset$.
Then, the fusion weights for the $g$-th cluster $\{ \omega_{i,g} \}_{{i \in \mathcal{N}}}$ can be computed as,
\begin{equation}\label{eq:Group FW}
\omega_{i,g}=\psi_{{i,g}}\left/{\sum}_{i \in \mathcal{N}} \psi_{{i,g}}\right.,
\end{equation}
where the denominator is a normalisation constant to ensure ${\sum}_{i\in\mathcal{N}}\omega_{i,g}=1$. The ${G_{\mathcal{C}}}$ sets of scalar fusion weights for all groups are denoted as $\mathcal{W} = \{ \{ \omega_{i,g}\}_{i \in \mathcal{N}} \}_{g=1}^{G_{\mathcal{C}}}$. By now, it is worth noting that the complex and space-varying relative information confidence embedded in the MODs of different sensors has been disintegrated and characterised by set $\mathcal{W}$. Next, based on $\mathcal{W}$, homogeneous information averaging can be properly performed within each cluster via any reasonable fusion rule. The fused ${G_{\mathcal{C}}}$ sub-MODs are denoted as $\{\bar{\pi}_{g}\}_{g=1}^{G_{\mathcal{C}}}$.

\begin{Rem}
The rationale behind sub-density level homogeneous fusion lies in three aspects: a) the sub-MODs are clustered into groups in a way that the information averaging within each group is feasible due to C.3; b) {the RFSs of the sub-MODs in different groups are mutually independent because of C.1 and hence the information fusion among independent RFSs across groups is not necessary;} c) although different groups have different {``local conditions''}, {the relative information confidence of the sub-MODs within a cluster can be sufficiently described by its associated weights due to C.2 and C.4, thus homogeneous information fusion within each cluster is reasonable.}
\end{Rem}

\vspace{3mm}
\underline{\noindent {3) Reconstruction of the fused joint MOD}}
\vspace{3mm}

Our final objective is to obtain a fused MOD which can properly combine the information from all the local MODs $\{\pi_i(X)\}_{i\in\mathcal{N}}$. {As the MOD decomposition satisfies the C.1, the RFSs in the fused sub-MODs $\{\bar{\pi}_{g}\}_{g=1}^{G_{\mathcal{C}}}$ are also independent among each other due to the inheritance of the independence.} Thus, based on the convolution formula~\cite[(11.252)]{Mahler2007Book}, the fused joint MOD, denoted by $\breve{\pi}(X)$, can be reconstructed, without information loss, using the fused sub-MODs $\{\bar{\pi}_{g}\}_{g=1}^{G_{\mathcal{C}}}$ as,
\begin{equation}\label{eq:convo_recons}
\breve{\pi}(X) = \sum\limits_{ X_1 \biguplus \ldots \biguplus X_{G_{\mathcal{C}}} = X } {\bar{\pi}}_{1}(X_1)\cdots{\bar{\pi}}_{{G_{\mathcal{C}}}}(X_{G_{\mathcal{C}}}),
\end{equation}
where $X = \cup_{g=1}^{G_{\mathcal{C}}} X_g$. Through (\ref{eq:convo_recons}), we can see that the obtained $\breve{\pi}(X)$ includes the complete information of all the local MODs, and it combines the information in a way that the different {``local conditions''} of the local MODs being properly accounted by $\mathcal{W}$.
\begin{Rem}
In general, the reconstructed joint MOD $\breve{\pi}(X)$ might not be of the same type of the averaged local MODs. In this case, we have to approximate $\breve{\pi}(X)$ by an MOD of the type under consideration to continue with the subsequent filtering recursion. The approximated MOD can be, for instance, the MOD which can preserves the first and second order moments of the original MOD \cite{Vo2009,Reuter2014}, or the MOD which can minimize the KLD with respect to the true MOD \cite{Battistelli2013,Giorgio_KLA}.
\end{Rem}

\subsection{Summary and Discussion}\label{sec:Summary}
As mentioned before, the key spirit of the proposed heterogeneous fusion method is the \emph{local condition tailored} tactic. By means of the MOD decomposition, the complex local MODs with space-varying information confidence are disintegrated into smaller size sub-MODs with homogeneous information confidence. Then, clustering is performed to transfer the original complicated fusion problem into a parallelizable multi-cluster fusion problem, wherein each inner-cluster fusion can be executed, simply and independently, based on the associated tailor-made fusion weights according to the existing homogeneous fusion strategies. {Note that if the WGA and WAA fusion rules are adopted for the information averaging within each clustered group, then all the inner-cluster homogeneous fusions are immune to the double counting of common information due to the unknown correlations of measurements~\cite{EMD-Julier2006,Bailey2012}. Hence, the heterogeneous fusion method is immune to the double counting issue as well.}

By now, there are two things that need to be noted. First, the conditions C.1--C.4, which are prerequisites for the heterogeneous fusion, may not hold ideally for some scenarios due to the complexities of the sensor characteristics, environmental factors and their coupling effect on the MODs. In these cases, direct implementation of heterogeneous fusion is infeasible and suboptimal approximations are needed. Second, how to quantitatively and systematically evaluate the ICC $\phi_{i,m}$ for each factorized sub-MOD is another crucial point, as it is the basis of the clustering process and the heterogeneous weight computation. Also, its evaluation process needs to carefully take into account both the sensor and environmental factors. We will illustrate clearly how C.1--C.4 are achieved and how $\phi_{i,m}$ is quantitatively evaluated in the subsequent Sections \ref{sec:PHDimplem} and \ref{sec:MBimplem}, wherein we derive two step-by-step heterogeneous fusion algorithms with the MPP and MB MODs.

\section{Analytic Implementations using PHD Filters}\label{sec:PHDimplem}
In Section \ref{sec:HFS}, the general framework of heterogeneous fusion strategy is introduced without any particular assumptions on the types of MODs. Here, we apply this strategy with the MPP MODs, and derive the analytic implementation equations for the heterogeneous multi-sensor PHD (H-MPHD) filter.

\subsection{Heterogeneous Multi-sensor PHD Filter}\label{sec:HPHD}

Consider a {centralized sensor network} in which PHD filters are running locally in every sensor node, and recursively propagating MPP MODs for multi-object tracking. Let $\{\pi^{\text{\texttt{P}}}_i(X)\}_{i\in\mathcal{N}}$ denote the set of the MPP MODs obtained by the local PHD filters. At each time, the local MPP MODs $\{\pi^{\text{\texttt{P}}}_i(X)\}_{i\in\mathcal{N}}$ are sent to a fusion center to compute a fused global MPP MOD according to the heterogeneous fusion strategy.

\vspace{3mm}
\underline{\noindent {1) Space partition based MOD factorization}}
\vspace{3mm}

According to Section \ref{sec:generalm}, the first step is to decompose the MPP MODs $\{\pi^{\text{\texttt{P}}}_i(X)\}_{i\in\mathcal{N}}$ into sub-MODs which can satisfy both C.1 and C.2. By definition \cite{Mahler2007Book,MahlerPHD,Vo2006}, all the elements of an MPP RFS are independently distributed, thus C.1 can naturally hold during the MOD factorization. As for C.2, we propose a space partition based MOD decomposition method. First, the definition of a space partition is given as follows.
\begin{Def}\label{Def:Space-partition}
Given a state space $\mathbb{X}$, we define an $M$-portion space partition of $\mathbb{X}$ as,
\begin{equation}\label{SpacePartition}
\mathcal{X}_M(\mathbb{X})=\left\{ \mathbb{X}_1, \ldots,\mathbb{X}_M \right\}
\end{equation}
with
\begin{align}
\mathbb{X}&=\cup_{m=1}^M\mathbb{X}_m,\hspace{2mm}{M\geq1},\\
\mathbb{X}_n\cap \mathbb{X}_m&=\emptyset, \hspace{2mm}n,m \in \{1, \ldots, M\},\hspace{1mm} n \neq m.
\end{align}
\end{Def}
We can see from Definition \ref{Def:Space-partition} that the partitioned $M$ subspaces $\left\{\mathbb{X}_m \right\}_{m=1}^M$ constitute the complete state space $\mathbb{X}$, and are mutually disjoint among each other.
\begin{Def}\label{Def:SpacePartiDecomp}
Given an $M$-portion space partition $\mathcal{X}_M(\mathbb{X})$ of the form (\ref{SpacePartition}), we define a space partition based MOD decomposition of $\pi({X})$ according to $\mathcal{X}_M(\mathbb{X})$ as $\{\pi_m({X})\}_{m=1}^M$, where
$\pi_m({X})=\pi({X}|X \subset\mathbb{X}_m)$.
\end{Def}
Following Definition \ref{Def:SpacePartiDecomp}, every local MPP MOD $\pi^{\text{\texttt{P}}}_i(X)$, $i\in\mathcal{N}$, can be decomposed into $M$ sub-MODs $\{\pi_{i,m}\}_{m=1}^{M}$, {and the MPP RFSs of these sub-MODs are mutually independent among each other.} In order to satisfy C.2, the space partition $\mathcal{X}_M(\mathbb{X})$ is performed in such a way that the information confidence within each sub-MOD $\pi_{i,m}$ is relatively consistent and can be quantified by a scalar-type ICC $\phi_{i,m}$. Later in Section \ref{sec:HPHD_Weights}, we will show how ICC $\phi_{i,m}$ can be computed systematically.

\begin{Pro}\label{Pro:PoiPartition}
Given an MPP MOD $\pi^{\text{\texttt{P}}}$ of the form (\ref{Poi distribution}) defined on state space ${\mathbb{X}}$, and a space partition based MOD decomposition $\{\pi_m\}_{m=1}^M$ of $\pi^{\text{\texttt{P}}}$ according to an $M$-portion space partition $\mathcal{X}_M(\mathbb{X})$ defined by \eqref{SpacePartition}, then the $m$-th partitioned sub-MOD is also an MPP MOD and of the following form,
\begin{equation}\label{part Poi}
{\pi^{\text{\texttt{P}}}_m}\left( X \right) = {e^{ - {\lambda _m}}}{\prod}_{x \in {X}} {\lambda _m}{f_m \left( x \right)},
\end{equation}
where
\begin{align}
{f_m}(x) &= \frac{{f\left( x \right)} {I_{\mathbb{X}_m}(x)} }{p_m},\\
{\lambda_m}&= {\lambda}{p_m},\\
{p_m} &= \int_{\mathbb{X}_m} {f\left( x \right)dx}.
\end{align}
and ${I_{S}(\cdot)}$ denotes the indicator function on the set $S$.
\end{Pro}
The proof of Proposition \ref{Pro:PoiPartition} is given in Appendix \ref{Prf:PoiPartition}. Proposition \ref{Pro:PoiPartition} shows that decomposed sub-MODs are also MPP distributed, thus denoted as $\{\pi^{\text{\texttt{P}}}_{i,m}\}_{m=1}^M$.

\vspace{3mm}
\underline{\noindent {2) Sub-density level fusion using the WAA fusion rule}}
\vspace{3mm}

As all the local MODs are factorized based on a same space partition $\mathcal{X}_M(\mathbb{X})$, the decomposed sub-MODs are automatically clustered into $M$ groups as $\{\{\pi^{\text{\texttt{P}}}_{i,m}\}_{i\in\mathcal{N}}\}_{m=1}^M$, and the information fusion among the sub-MODs within each group is naturally feasible. Therefore, C.3 and C.4 hold readily.
The heterogeneous fusion weights $\{\{\omega_{i,m}\}_{i\in\mathcal{N}}\}_{m=1}^M$ can also be directly computed using ICCs $\{\{\psi_{i,m}\}_{i\in\mathcal{N}}\}_{m=1}^M$ according to \eqref{eq:Group FW}. By now, using these tailor-made weights, information averaging can be performed appropriately within each group according to a certain homogeneous fusion rule.

Here, the WAA fusion rule is chosen as the illustration example due to its cardinality robustness compared to the WGA~\cite{Ueny2019_TAES,LTC2019}. Recently, the analytical expressions of the WAA fusion have been derived for MPP MOD in~\cite{Reza2017,gao2020multiobject,Kai2019} with different theoretical interpretations. Let $\{{\pi_{i,m}^{\text{\texttt{P}}}}\}_{i \in \mathcal{N}}$ and $\{ \omega_{i,m}\}_{i \in \mathcal{N}}$ denote, respectively, the sub-MODs and the fusion weights of the $m$-th cluster, then the fused density ${\bar{\pi}_{m}^{\text{\texttt{P}}}}$ according to WAA is also an MPP MOD with the following expression,
\begin{equation}\label{eq:part Poi WAA}
\bar{\pi}^{\text{\texttt{P}}}_m(X) = e^{ - \bar{\lambda}_m}{\prod}_{x\in X} \bar{\lambda}_m \bar{f}_m(x)
\end{equation}
where
\begin{align}
\label{eq:part Poi WAA1}
\bar{\lambda}_m &= {\sum}_{i\in\mathcal{N}} \omega_{i,m} \lambda_{i,m}, \\
\label{eq:part Poi WAA2}
\bar{f}_m(x) &= \frac{ \sum_{i\in\mathcal{N}} \omega_{i,m} \lambda_{i,m} f_{i,m}(x) }{ \sum_{i\in \mathcal{N}} \omega_{i,m} \lambda_{i,m} }.
\end{align}
{In principle, other fusion rules such as the WGA fusion rule can also be adopted but is out of the scope of this work.}

\vspace{3mm}
\underline{\noindent {3) Reconstruction of the fused joint MOD}}
\vspace{3mm}

{As the MPP RFSs of the fused sub-MODs $\{\bar{\pi}^{\text{\texttt{P}}}_m \}_{m=1}^M$ are mutually independent,} the fused joint MOD $\breve{\pi}$ can be constructed using the convolution formula~\cite{Mahler2007Book} as that in (\ref{eq:convo_recons}). Before proceeding further, the following Proposition \ref{Pro:Poi_Union} is given first.
\begin{Pro}\label{Pro:Poi_Union}
Suppose that an RFS $X = \cup_{m=1}^M X_m$ is the union of $M$ statistically independent MPP subsets $\{X_m\}_{m=1}^M$ with their MODs denoted as $\{{\pi}^{\text{\texttt{P}}}_m \}_{m=1}^M$ and parameterized by $\{(\lambda_m,f_m)\}_{m=1}^M$, then the set union $X$ is also an MPP RFS with its MOD given by,
\begin{equation}\label{eq:joint Poi}
\breve\pi^{\text{\texttt{P}}}(X) = e^{ -  \breve{\lambda} }{\prod}_{x \in X}  \breve{\lambda} \cdot  \breve f(x)
\end{equation}
where
\begin{align}
 \breve{\lambda}  &= {\sum}_{m = 1}^M {\lambda}_m, \\
\breve{ f}(x) &= {\sum}_{m = 1}^M  \frac{{\lambda}_m}{ \breve{\lambda}} {f}_m(x).
\end{align}
\end{Pro}
The detailed proof of Proposition \ref{Pro:Poi_Union} can be found in Appendix \ref{Prf:Poi_Union}. Proposition \ref{Pro:Poi_Union} establishes the statistical relationship between the independent MPP subsets and their union. Substitution of \eqref{eq:part Poi WAA}--\eqref{eq:part Poi WAA2} into Proposition \ref{Pro:Poi_Union} yields the following Proposition \ref{Pro:H-MS-PHD}, which summarizes the detailed implementation equations of the proposed H-MPHD filter.
\begin{Pro}\label{Pro:H-MS-PHD}
Let $\nu^{\text{\texttt{P}}}_i(x)$, $\lambda_i$ and $f_i(x)$ denote, respectively, the intensity function, mean target number and location density of the $i$-th, ${i \in \mathcal{N}}$, local PHD filter. Given an $M$-portion space partition $\mathcal{X}_M(\mathbb{X})$ of the form (\ref{SpacePartition}) and the corresponding heterogeneous fusion weights $\mathcal{W} = \{ \{ \omega_{i,m}\}_{i \in \mathcal{N}} \}_{m=1}^M$, then the fused MOD by H-MPHD filter is an MPP MOD with its intensity function $\breve{\nu}^{\text{\texttt{P}}}(x)$ and mean target number $\breve{\lambda}$ given by,
\begin{align}
\label{eq:H-MS-PHD_phd}
\breve{\nu}^{\text{\texttt{P}}}(x)  &= {\sum}_{i\in\mathcal{N}} {\nu}^{\text{\texttt{P}}}_i(x) \omega_i(x), \\
\label{eq:H-MS-PHD_mn}
 \breve{\lambda}  &= {\sum}_{i\in\mathcal{N}}{\lambda}_i \mu_i
\end{align}
where
\begin{align}
\label{eq:H-MS-PHD_omega}
\omega_i(x)  &=  {\sum}_{m=1}^M \omega_{i,m} I_{\mathbb{X}_m}(x), \\
\label{eq:H-MS-PHD_mu}
\mu_i&={\sum}_{m = 1}^M \omega_{i,m}\hspace{1mm}p_{i,m},\\
\label{eq:H-MS-PHD_p}
p_{i,m} &= \int_{\mathbb{X}_m} {f_i(x)dx}.
\end{align}
\end{Pro}
\begin{Cor}\label{Cor:Poi_Union}
Suppose that the heterogeneous weights are set homogeneously as $\omega_{i,m}=\omega_i$ for all the factorized MPP sub-MODs $\{\pi^{\text{\texttt{P}}}_{i,m}\}_{m=1}^M$, then equations \eqref{eq:H-MS-PHD_phd} and \eqref{eq:H-MS-PHD_mn} become $\breve{\nu}^{\text{\texttt{P}}}(x)= {\sum}_{i\in\mathcal{N}} {\nu}^{\text{\texttt{P}}}_i(x) \omega_i$ and $ \breve{\lambda}= \sum_{i\in\mathcal{N}}{\lambda}_i \omega_i$, respectively.
\end{Cor}
We can see from \eqref{eq:H-MS-PHD_phd} and \eqref{eq:H-MS-PHD_mn} that the H-MPHD filter adopts a simple and intuitive structure.
Similar to the homogeneous WAA, the fused intensity function $\breve{\nu}^{\text{\texttt{P}}}(x)$ and expected target number $\breve{\lambda}$ are also the weighted summations of the corresponding local statistics. Corollary \ref{Cor:Poi_Union} further shows that H-MPHD filter becomes exactly a standard WAA based multi-sensor PHD filter when the weights are homogeneously set.

However, it can be seen from \eqref{eq:H-MS-PHD_omega} that the key difference of the H-MPHD filter is that the weight $\omega_i(x)$ for intensity summation are no longer a fixed scalar but a state-dependent function with a terrace shape. For any state $x \in \mathbb{X}$, {the corresponding weights $\{{{\omega}_{i}(x)} \}_{i \in \mathcal{N}}$ sum to one}, and reflect the relative information confidence of the local intensity functions on that state. {Likewise, equations \eqref{eq:H-MS-PHD_mu} and \eqref{eq:H-MS-PHD_p} show that the weights $\{\mu_i\}_{i \in \mathcal{N}}$ for target number summation also depend on the heterogeneous fusion weights $\mathcal{W}$. Specifically, in order to quantify the contribution of each local MOD on the mean target number $\breve{\lambda}$, the term $\mu_i$ is also a heterogeneously weighted average considering the space-varying relative information confidence of the $i$-th MOD $\pi^{\text{\texttt{P}}}_i$ on the partitioned $M$ subspaces $\left\{\mathbb{X}_m \right\}_{m=1}^M$.} In this way, the H-MPHD filter is capable of performing information averaging of MODs in a more flexible manner. Mathematically, the knowledge of sensor sensing abilities and environment inputs can be properly embodied into the fusion process by dynamically adjusting $\{{{\omega}_{i}(x)}\}_{i  \in \mathcal{N}}$ and $\{\mu_i\}_{i \in \mathcal{N}}$ through $\mathcal{W}$.
\begin{Rem}
Proposition \ref{Pro:H-MS-PHD} shows that the reconstructed $\breve{\pi}(X)$  is exactly an MPP MOD, which allows the subsequent filtering recursion to be performed readily using $\breve{\pi}(X)$ as a prior MOD. Hence, the proposed H-MPHD filter enjoys a conjugate closure which is highly desirable in Bayesian inference~\cite{Vo2013}.
\end{Rem}
\begin{Rem}\label{Rem:HMPHDC14}
It is important to note that the prerequisites C.1, C.3 and C.4 directly hold thanks to the independence property of MPP and the space partition based MOD decomposition. Also, in principle, if the state space $\mathbb{X}$ is partitioned into tiny subsets, the relative information confidence for each subset is constant without regard for the specific characteristics of sensor abilities and environmental factors. In this case, C.2 is satisfied. That means the implementation of the H-MPHD filter is feasible in general cases without the need for any suboptimal approximations.
\end{Rem}
\begin{Rem}
In principle, a heterogeneous multi-sensor CPHD filter can be devised similarly by applying a space partition based factorization of i.i.d. cluster process MOD, then reconstructing the joint MOD after sub-MOD level fusion. However, the reconstructed MOD need not to be an i.i.d. cluster process MOD. In this case, both accurate approximation and efficient implementation are required, which is not trivial and out of the scope of this work.
\end{Rem}

\subsection{Design of the Heterogeneous Fusion Weights}\label{sec:HPHD_Weights}
Here we show how the heterogeneous fusion weights $\mathcal{W}$ can be evaluated in a principled way. For any object with state $x\in\mathbb{X}$, its estimation uncertainty is mainly decided by the quality of its associated measurements and the employed estimation approach, e.g., the type of chosen filter. Let $\Xi(x)\subseteq \mathbb{Z}$ denote the set of measurements which are closely related to the estimation of $x$, and its specific form depends on how a subject object can affect the sensor measurements. Further, let $e$ and $s$ denote, respectively, the parameters that quantitatively characterize the sensor ability and environmental factors. {Further, let $e=\{e^{(r)}\}_{r=1}^{n_e}$ and $s=\{s^{(l)}\}_{l=1}^{n_s}$ denote, respectively, the parameters that quantitatively characterize the environmental factors and sensor abilities. The integers ${n_e}$ and ${n_s}$ are the numbers of parameters in the two sets. These parameters can be, for example, sensor resolution, covariance matrix of measurement errors, clutter intensity, positions of sight obstructions and so on.} Clearly, the quality of the measurements in $\Xi(x)$ is statistically related to {sets $e$ and $s$}. In the following, we first define an estimation uncertainty function (EUF) in order to quantify the estimation uncertainty of any object with state $x\in\mathbb{X}$ under the environment and sensor conditions parameterized by sets $e$ and $s$.
\begin{Def}\label{Def:EUF}
Let $e$ and $s$ denote, respectively, the parameter sets of environmental inputs and sensor abilities. Then, the EUF is defined on $\mathbb{X}$ and of the following form,
\begin{equation}\label{eq:EUF}
\Phi(x;e,s) \triangleq \Phi^a(x;e,s)\cdot u_1 + \Phi^c(x;e,s) \cdot u_2,
\end{equation}
where $\Phi^a$ and $\Phi^c$ are the accuracy EUF (AEUF) and cardinality EUF (CEUF), respectively. The non-negative scalars $u_1$ and $u_2$ are used to transfer cardinality and accuracy uncertainties into a unitless total uncertainty coefficient.
\end{Def}
It can be seen from \eqref{eq:EUF} that the EUF $\Phi(x;e,s)$ takes into consideration both sensor and environment impacts. The AEUF $\Phi^a$ and CEUF $\Phi^c$ quantify the localization and cardinality estimation uncertainties, respectively. The two scalar coefficients $u_1$ and $u_2$ are used to adjust the relative penalties of the two types of uncertainties and can be user-defined according to specific application requirements. {In this way, the EUF $\Phi(x;e,s)$ is able to jointly capture both accuracy and cardinality uncertainties. The structure of EUF is inspired by the optimal sub-pattern assignment (OSPA) distance~\cite{Schumacher}, which is designed to jointly capture the localization and cardinality errors during multi-object estimation. The scalar coefficients $u_1$ and $u_2$ in \eqref{eq:EUF} are inspired by the cut-off value $c$ in OSPA, which is employed to transfer the cardinality error into a total localization error. Further details about the OSPA distance can be found in\cite{Schumacher}.}

The design of AEUF $\Phi^a$ and CEUF $\Phi^c$ is an open problem. In the following, we provide a simple but reasonable example.
First, the estimation covariance $P(x)$ of a state $x$ is often used to characterize its localization estimation uncertainty, and $P(x)$ has already combined the impacts from sensor, environment and also estimation approach. Therefore, a natural and possible choice of AEUF $\Phi^a$ is the determinant of $P(x)$,
\begin{equation}\label{eq:AEUF1}
\Phi^a(x;e,s)\triangleq |P(x)|_\text{d},
\end{equation}
where $|\cdot|_\text{d}$ denotes the determinant operation. It has also been proved that $|P(x)|_\text{d}$ is effective in the weight selection of CI fusion rule~\cite{Hurley2002}. However, in general cases, the covariance $P(x)$ cannot be available for all the states. Therefore, a feasible alternative is needed. We recall that the estimation uncertainty of $x$ is heavily related to the quality of its associated measurements $\Xi(x)$. For any $z\in \mathbb{Z}$, let $R^x(z)$ denote its coordinate-converted measurement covariance matrix. {The derivation of $R^x(z)$ can be found in ~\cite[(13a--13c)]{lerro1993tracking}, and the detailed expressions of $R^x(z)$ for the Cartesian to polar coordinates conversion are given in the Appendix C for the reader's convenience.} Then, a possible substitute of $P(x)$ is the weighted average of the covariance matrixes of the measurements in set $\Xi(x)$ based on its extent of relevance to the estimation of state $x$, and is denoted by $\bar{R}^x(\Xi)$. It is worth noting that $\bar{R}^x(\Xi)$ is not used to approximate $P(x)$, but rather an alternative to reflect the relative estimation uncertainty among object states.

The cardinality estimation uncertainty can be affected by many factors, for example, the decline of detection probability due to sensor range attenuation, intensity of the clutter measurements, limited sensor FoV, sight obstruction and so on. In order to take these factors into consideration, a reasonable form of CEUF $\Phi^c$ is given as follows,
\begin{equation}\label{eq:CEUF1}
\Phi^c(x;e,s)\triangleq \frac{\lambda_c^x(\Xi)}{P_D(x)\digamma(x)}.
\end{equation}
where $P_D(x)$ denotes the detection probability of object with state $x$, $\lambda_c^x(\Xi)$ represents the intensity of the false measurements that can be associated with $x$, and $\digamma(x)=1$ if the object $x$ is within the sensor FoV and not blocked by any sight obstruction, and otherwise $\digamma(x)=0$. Note that other factors can also be taken into account by integrating corresponding parameters into $\Phi^c$. The substitution of \eqref{eq:AEUF1} and \eqref{eq:CEUF1} into \eqref{eq:EUF} then yields a computable EUF.

Next, we show how the ICCs and heterogeneous weights $\mathcal{W}$ can be evaluated using EUF $\Phi(x;e,s)$. By definition, an MPP MOD $\pi^{\text{\texttt{P}}}$ is completely characterized by its intensity function $\nu^{\text{\texttt{P}}}(x)$, certainly including its information confidence. Intuitively, the quality of the information embedded in $\nu^{\text{\texttt{P}}}(x)$ degrades as the estimation uncertainty increases. Therefore, as for any factorized sub-MOD $\pi^{\text{\texttt{P}}}_{i,m}$, which is defined on the $m$-th partition sub-space $\mathbb{X}_m$, its ICC can be computed as,
\begin{equation}\label{eq:ICC_PHD1}
\phi_{i,m}=\left[\Phi_i(x_m^\ast;e_i,s_i)\right]^{-1},
\end{equation}
where $\Phi_i$ denotes the EUF of the $i$-th sensor, and $x_m^\ast$ can be any state variable in $\mathbb{X}_m$ when C.2 is satisfied. Then, by substituting \eqref{eq:ICC_PHD1} into \eqref{eq:Group FW}, the heterogeneous weights $\mathcal{W}$ can be obtained. Note that, if the state space $\mathbb{X}$ is partitioned into {tiny subsets}, the $\omega_i(x)$ of \eqref{eq:H-MS-PHD_omega} can be directly computed as,
\begin{equation}\label{eq:omega_PHD1}
\omega_i(x)=\frac{\left[\Phi_i(x;e_i,s_i)\right]^{-1}}{\sum\limits_{j\in\mathcal{N}}\left[\Phi_j(x;e_j,s_j)\right]^{-1}}.
\end{equation}
\begin{Rem}
The defined EUF of \eqref{eq:EUF} is not intend to provide a rigorous and absolute calculation or quantification of the estimation uncertainty. Rather, it is an intermediate quantity to evaluate the relative information confidence among MODs.
\end{Rem}

\subsection{Gaussian Mixture Implementation}\label{sec:GM-HPHD}
Compared with SMC implementation, the GM approach is only suitable for the Gaussian and linear or weak non-linear cases~\cite{Vo2006}, but it can facilitate the state estimation process, and usually has lower transmission and computation load. Here we briefly discuss the GM implementation of the proposed H-MPHD filter. Suppose that the $i$-th local intensity function $\nu_i^{\text{\texttt{P}}}(x)$, ${i \in \mathcal{N}}$, is represented as a mixture of ${J_{i}}$ Gaussian components with weights $\{\alpha_{i}^{j}\}_{j=1}^{J_{i}}$ as
\begin{equation}
\begin{split}\label{eq:GMs}
{\nu}^{\text{\texttt{P}}}_{i}(x)={\sum}_{j=1}^{J_{i}}\alpha_{i}^{j}\mathcal{G}\left(x;m_{i}^{j},P_{i}^{j}\right),
\end{split}
\end{equation}
where $\mathcal{G}\left(x;m,P\right)$ denotes a Gaussian density with mean $m$ and covariance $P$. Then, the fused intensity function outputted by H-MPHD filter can be obtain by substituting (\ref{eq:GMs}) into (\ref{eq:H-MS-PHD_phd}),
\begin{equation}\label{eq:H-MPHD_GM}
\breve{\nu}^{\text{\texttt{P}}}(x)= \sum\limits_{i  \in \mathcal{N}}  {{{\omega} _i}(x) } \sum\limits_{j=1}^{J_{i}}\alpha_{i}^{j}\mathcal{G}\left(x;m_{i}^{j},P_{i}^{j}\right).
\end{equation}
Due to the modulation of ${{{\omega} _i}(x) }$, (\ref{eq:H-MPHD_GM}) is no longer a standard mixture of Gaussian components. In order to facilitate subsequent processing, e.g., state extraction, one possible solution is to approximate (\ref{eq:H-MPHD_GM}) as follows to retain the GM structure,
\begin{equation}\label{eq:H-MPHD_GM1}
\breve{\nu}^{\text{\texttt{P}}}(x)\approx\sum\limits_{i \in \mathcal{N}} \sum\limits_{j=1}^{J_{i}} {{\omega}_{i}(m_{i}^{j})} \alpha_{i}^{j}\mathcal{G}\left(x;m_{i}^{j},P_{i}^{j}\right).
\end{equation}
Similar approximation has been employed and tested in \cite{Yi2020TSPDFoVs}. In this case, the expected target number can be computed as,
\begin{equation}\label{eq:H-MPHD_GM2}
\breve{\lambda}={\sum}_{j=1}^{J_{i}}{{\omega}_{i}(m_{i}^{j})}.
\end{equation}
Using \eqref{eq:H-MPHD_GM1} and \eqref{eq:H-MPHD_GM2}, multi-object states can be readily extracted by rounding $\breve{\lambda}$ into $\lfloor\breve{\lambda}\rfloor$, {where $\lfloor{\cdot}\rfloor$ denotes the floor operator,} and then selecting the $\lfloor\breve{\lambda}\rfloor$ highest peaks from $\breve{\nu}^{\text{\texttt{P}}}(x)$~\cite{Vo2006}. Alternative approximation strategy can be, for instance, finding the best GM representation of (\ref{eq:H-MPHD_GM}) according to a certain criteria, but is not considered here.

Another issue is that the number of Gaussian components increases linearly after the fusion of \eqref{eq:H-MPHD_GM1}. In order to control the component number, classic strategies such as pruning and merging of Gaussian components \cite{Vo2006} need to be applied.

{\subsection{Algorithm Complexity}\label{sec:GM-HPHD-Cost}
It can be seen from \eqref{eq:H-MPHD_GM1} that the fusion complexity of the H-MPHD filter mainly contains two parts. Firstly, we need to assign the heterogeneous fusion weights to all Gaussian components, and the corresponding complexity is about $\mathcal{O}({\sum}_{i\in\mathcal{N}}{J_{i}})$. Then, the weighted summation of all Gaussian components also results a complexity about $\mathcal{O}({\sum}_{i\in\mathcal{N}}{J_{i}})$, and hence the total fusion complexity is about $\mathcal{O}(2{\sum}_{i\in\mathcal{N}}{J_{i}})$. It should be noted that we do not take into account the complexity for computing the heterogeneous fusion weights, which has a complexity of $\mathcal{O}(M)$ with $M$ the number of partitioned sub-MODs. The reason is that these weights can be computed offline and stored in the fusion center. There is no need to repeatedly compute and transmit the fusion weights as the sensor abilities and environmental factors do not change at each fusion interval.
Even if the sensor abilities and environmental factors are changed, each sensor can update the its corresponding EUF locally
and only send the updated EUF to the fusion center for the calculation of the heterogeneous fusion weights. In comparison, the standard WAA fusion does not have the complexity for assigning the fusion weights since all Gaussian components of a local intensity function adopt a same fusion weight.}

\section{Analytic Implementations using MB Filters}\label{sec:MBimplem}

In this section, we discuss how the proposed heterogeneous fusion strategy is implemented in a {centralized sensor network} where each sensor node runs an MB filter.

\subsection{Heterogeneous Multi-sensor MB Filter}\label{sec:HMB}

Similarly, the development of the corresponding heterogeneous multi-sensor MB (H-MMB) filter follows the three major steps as that in Sections \ref{sec:generalm} and \ref{sec:HPHD}. {The aim of the H-MMB filter is to properly integrated all the local MB MODs into a fused global MB MOD at a fusion center.}

\vspace{3mm}
\underline{\noindent {1) Object-oriented MOD factorization}}
\vspace{3mm}

By definition~\cite{Mahler2007Book}, an MB RFS is a union of a fixed number of independent Bernoulli RFSs. The key spirit of MB filter for multi-object estimation is its component-to-object pairing estimation strategy. Namely, each Bernoulli component is used to model the statistical uncertainties of a specific hypothesized object. Let $\pi^{\text{\texttt{MB}}}_i$ denote the MB MOD propagated by the $i$-th, $i\in\mathcal{N}$, local MB filter, and it is parameterized as,
\begin{equation}
\pi^{\text{\texttt{MB}}}_i=\left\{\left(r_{i,b},f_{i,b}\right)\right\}_{b\in \mathbb{B}_{i}},
\end{equation}
where $\mathbb{B}_{i}\triangleq\{1,\cdots,B_{i}\}$ is the set of indexes of the constituent Bernoulli components. The terms $r_{i,b}$ and $f_{i,b}(\cdot)$ describe, respectively, the existence uncertainty and state distribution of the potential object modeled by the $b$-th Bernoulli component.

In this case, the space partition based MOD factorization adopted in Section \ref{sec:HPHD} is no longer suitable herein since it does not well exploit the component-constituent structure of the MB MODs. A natural and physically meaningful choice is to apply an object-oriented MOD factorization. Specifically, the $i$-th MB MOD $\pi^{\text{\texttt{MB}}}_i$ is factorized into a set of $B_{i}$ Bernoulli sub-density $\{\pi^{\text{\texttt{B}}}_{i,b}\}_{b\in \mathbb{B}_{i}}$, where $\pi^{\text{\texttt{B}}}_{i,b}$ denotes a Bernoulli density parameterized by $r_{i,b}$ and $f_{i,b}(\cdot)$. Note that condition C.1 holds when the object-oriented MOD factorization is applied, since, by definition, all the Bernoulli components in an MB RFS are mutually independent.

With regard to the validation of C.2, we first give the definition of the HPD region~\cite{HPD2010} as follows.
\begin{Def}\label{Def:HPD_Def}
Let $f(x)$ be a probability density function defined on state space $\mathbb{X}$. If a region $S_\alpha(f)\subset \mathbb{X}$ satisfies
\begin{itemize}
  \item $\Pr \{ x \in S_\alpha(f)\}  = \alpha,\hspace{4mm}0<\alpha < 1,$
  \item $p(x_1)\geq p(x_2), \hspace{4mm}x_1 \in S_\alpha(f); x_2 \notin S_\alpha(f),$
\end{itemize}
where $\alpha$ is a positive scalar close to one, e.g., $\alpha=0.95$, then $S_\alpha(f)$ is called an HPD $\alpha$-credible region of $f(x)$.	
\end{Def}
Thus, the validation of C.2 holds if the location density $f_{i,b}(\cdot)$ of any factorized Bernoulli component $\pi^{\text{\texttt{B}}}_{i,b}$ satisfies,
\begin{equation}\label{eq:HMMB_C2}
|\Phi(x_1;e,s)-\Phi(x_2;e,s)|\leq\Delta_\varepsilon
\end{equation}
where $\Phi(\cdot;e,s)$ is the EUF defined in \eqref{eq:EUF}, $x_1, x_2\in S_\alpha(f_{i,b})$, $x_1\neq x_2$, and $\Delta_\varepsilon$ is a threshold to judge the information consistency in the HPD region of $f_{i,b}(\cdot)$. We note that, usually, the HPD region of $f_{i,b}(\cdot)$ is small and \eqref{eq:HMMB_C2} holds since $f_{i,b}(\cdot)$ describes the spatial uncertainty of only a single object.

\vspace{3mm}
\underline{\noindent {2) Object-oriented fusion using the WAA fusion rule}}
\vspace{3mm}

Given the factorized Bernoulli densities $\{\pi^{\text{\texttt{B}}}_{i,b}\}_{b\in \mathbb{B}_{i}}$, ${i\in\mathcal{N}}$, {the next step is to cluster them into groups such that conditions C.3 and C.4 can be satisfied.} As mentioned above, the MB MODs are decomposed in a physically meaningful manner wherein each factorized Bernoulli density corresponds to a single hypothesized object. Therefore, the information averaging across multiple Bernoulli densities corresponding to different objects is physically infeasible and violating the spirit of the MB MOD. In the following, based on Definition \ref{clustering}, an object-oriented clustering is proposed to group the factorized Bernoulli densities.
\begin{Def}\label{Object-oriented-clustering}
Consider a clustering $C=\{\mathcal{C}_1,\cdots,\mathcal{C}_{G_{\mathcal{C}}}\}$ of Bernoulli components having index sets $\left\{  \mathbb{B}_i, i  \in \mathcal{N} \right\}$, where the $g$-th, $g=1,\ldots,G_{\mathcal{C}}$, cluster is denoted as $\mathcal{C}_{g}=\left\{\mathbb{M}_{i,g}, {i  \in \mathcal{N}}\right\}$, then $C$ is said to be an object-oriented clustering if any subset $\mathbb{M}_{i,g}$ is either a singleton or emptyset, and for any non-empty $\mathbb{M}_{i,g}$, its indexed Bernoulli density $\pi^{\text{\texttt{B}}}_{i,b}$, ${b\in \mathbb{M}_{i,g}}$, satisfies the following two conditions:
\begin{equation}\label{c-criterion1}
\min_{{{\scriptsize{\begin{array}{c}
i' \in \mathcal{N}, i'\neq i   \\
\mathbb{M}_{i',g}\neq\emptyset, b'\in \mathbb{M}_{i',g}
\end{array}}}}
} D_{\text{\texttt{KL}}}(\pi^{\text{\texttt{B}}}_{i,b}\|\pi^{\text{\texttt{B}}}_{i',b'})\leq\gamma,
\end{equation}
\begin{equation}\label{c-criterion2}
b= \arg\min_{l \in \mathbb{B}_{i} } \sum_{{{\scriptsize{\begin{array}{c}
i' \in \mathcal{N}, i'\neq i   \\
 b'\in \mathbb{M}_{i',g}
\end{array}}}}
} D_{\text{\texttt{KL}}}(\pi^{\text{\texttt{B}}}_{i,l}\|\pi^{\text{\texttt{B}}}_{i',b'}),
\end{equation}
where $\gamma$ is a predefined threshold.
\end{Def}
The inequality \eqref{c-criterion1} guarantees that, for any cluster $\mathcal{C}_{g}$, $g=1,\ldots,G_{\mathcal{C}}$, the contained Bernoulli components from different sensors are corresponding to a same potential object from the perspective of information consistency. Equation \eqref{c-criterion2} further demands that $\pi^{\text{\texttt{B}}}_{i,b}$, ${b\in \mathbb{M}_{i,g}}$, is the best Bernoulli density of sensor $i$, which matches with the potential object modelled by the $g$-th cluster. Therefore, if the factorized Bernoulli components can be grouped into an object-oriented clustering, C.3 and C.4 can be satisfied. Also, the information averaging within each cluster is feasible and also physically complying with the spirit of the MB MOD.

Regarding the construction of the object-oriented clustering, one efficient solution is to use the \emph{union find set} algorithm \cite{Geraud2005,Tarjan1975}, which only needs the pairwise KLD between Bernoulli components among sensors as inputs. In this way, the computational complexity of clustering is only polynomial in the number of Bernoulli components. Implementation details of this algorithm can be found in \cite{Yi2020TSP,Yi2020TSPDFoVs} wherein it has been used to cluster Bernoulli and Gaussian components. It should be noted that, after implementing \emph{union find set}, a subset $\mathbb{M}_{i,g}$ can contain {more than one Bernoulli component} when potential objects are in close proximity. In this case, one can sequentially examine the Bernoulli components in subset $\mathbb{M}_{i,g}$ and retain only the best one according to (\ref{c-criterion2}).

After clustering, we also choose the WAA fusion rule to implement the inner-cluster information averaging. The analytical fusion equations of Bernoulli densities have been derived recently in~\cite{li2019distributed,gao2020multiobject}. Let $\{{\pi_{i,b}^{\text{\texttt{B}}}},{b\in \mathbb{M}_{i,g}}\}_{i \in \mathcal{N}}$ and $\{ \omega_{i,g}\}_{i \in \mathcal{N}}$ denote, respectively, the factorized Bernoulli densities and the heterogeneous fusion weights of the $g$-th cluster of an object-oriented clustering, then the fused density according to WAA is also a Bernoulli density ${\bar{\pi}_{g}^{\text{\texttt{B}}}}$ parameterized by the following existence probability and spatial distribution~\cite{li2019distributed},
\begin{align}
\label{eq:part fusion Bernoulli1}
\bar r_g &= \sum\limits_{i\in\mathcal{N}, {b\in \mathbb{M}_{i,g}}} \omega_{i,g} r_{i,b},\\
\label{eq:part fusion Bernoulli2}
\bar f_g(x) &= \frac{1}{\bar r_g} \sum\limits_{i\in\mathcal{N}, {b\in \mathbb{M}_{i,g}}} \omega_{i,g} r_{i,b} f_{i,b}(x).
\end{align}
The evaluation of the heterogeneous fusion weights $\{ \omega_{i,g}\}_{i \in \mathcal{N}}$ will be discussed in detail in Section \ref{sec:HMMB_Weights}.
\begin{Rem}
In the multi-sensor fusion of the MB or labeled MB MODs, the component-pairing strategy has been adopted with the purposes of reducing the algorithm complexity \cite{Yi2020TSP}, dealing with the label inconsistency issues \cite{Suqi2018,Suqi2019}, and maintaining the closure of fusion equations \cite{li2020arithmetic}. Here, we advocate an alternative view of the use of the component-pairing strategy which aims to facilitate a local condition tailored information averaging.
\end{Rem}

\vspace{3mm}
\underline{\noindent {3) Reconstruction of the fused MB MOD}}
\vspace{3mm}

{The Bernoulli components of the fused densities $\{{\bar{\pi}_{g}^{\text{\texttt{B}}}}\}_{g=1}^{G_{\mathcal{C}}}$ are also independent among each other due to the inheritance of the independence of the factorized Bernoulli components.} Therefore, using the convolution formula~\cite{Mahler2007Book}, we can reconstruct the fused global MOD. The following Proposition \ref{Pro:H-MS-MB} summarizes the proposed H-MMB filter and shows that the reconstructed MOD $\breve{\pi}(X)$ is exactly an MB MOD consists of ${G_{\mathcal{C}}}$ Bernoulli components.
\begin{Pro}\label{Pro:H-MS-MB}
Let $\pi^{\text{\texttt{MB}}}_i=\{(r_{i,b},f_{i,b})\}_{b\in \mathbb{B}_{i}}$ denote the MB MOD of the $i$-th, ${i \in \mathcal{N}}$, local MB filter. Given an object oriented clustering $C=\{\mathcal{C}_1,\cdots,\mathcal{C}_{G_{\mathcal{C}}}\}$ of the factorized Bernoulli components of all the local MB MODs and the corresponding heterogeneous fusion weights $\mathcal{W} = \{ \{ \omega_{i,g}\}_{i \in \mathcal{N}} \}_{g=1}^{G_{\mathcal{C}}}$, then the fused MOD according to heterogeneous WAA fusion rule is an MB MOD parameterized by
\begin{equation}\label{eq:JointMB}
\breve{\pi}^{\text{\texttt{MB}}}=\left\{\left(\bar r_g,\bar f_g(x)\right)\right\}_{g=1}^{G_{\mathcal{C}}},
\end{equation}
where $\bar r_g$ and $\bar f_g(x)$ are given by \eqref{eq:part fusion Bernoulli1} and \eqref{eq:part fusion Bernoulli2}, respectively.
\end{Pro}

Proposition \ref{Pro:H-MS-MB} shows that the H-MMB filter also possesses a conjugate closure by maintaining the fused MOD an MB MOD. The fusion process of the H-MMB filter is performed within each cluster of the factorized Bernoulli components, in a parallelized manner, based on the tailor-made fusion weights. Also, it can be seen clearly in \eqref{eq:part fusion Bernoulli1} and \eqref{eq:part fusion Bernoulli2} that the knowledge of sensor abilities and environment inputs are embodied into the fusion process by affecting the computation of $\bar r_g$ and $\bar f_g(x)$ through $\mathcal{W}$.
Further, using the property of the MB MOD \cite[(16.66)]{Mahler2007Book}, the PHD of the fused MB MOD is given by,
\begin{align}\label{eq:HMMB-PHD}
\breve{\nu}^{\text{\texttt{MB}}}(x)= \sum\limits_{g=1}^{G_{\mathcal{C}}} \bar r_g \bar f_g(x)=\sum\limits_{g=1}^{G_{\mathcal{C}}}\sum\limits_{i\in\mathcal{N}, {b\in \mathbb{M}_{i,g}}} \omega_{i,g} r_{i,b} f_{i,b}(x).
\end{align}
\begin{Cor}\label{Cor:MB-PHD}
Suppose that the heterogeneous weights are set homogeneously as $\omega_{i,g}=\omega_i$ for all the factorized Bernoulli densities $\{\pi^{\text{\texttt{B}}}_{i,b}\}_{b\in \mathbb{B}_{i}}$, then $\breve{\pi}^{\text{\texttt{MB}}}$ defined by (\ref{eq:JointMB}) is the MB MOD that the matches exactly the first order moment of the MOD $\breve{\pi}$, which is the straightforward WAA fusion of the local MB densities defined as $\breve{\pi}={\sum}_{i\in {\mathcal{N}}} \omega_i \pi^{\text{\texttt{MB}}}_i$.
\end{Cor}
The proof of Corollary \ref{Cor:MB-PHD} can be found in Appendix \ref{Prf:Cor_MB-PHD}. Note that the direct WAA fusion of the local MB MODs does not result another MB MOD\cite{li2020arithmetic,gao2020multiobject}, and, therefore, approximation is needed to maintain conjugacy. Corollary \ref{Cor:MB-PHD} shows that, when homogeneous weights are employed, the fused MOD by the H-MMB filter is a reasonable approximation of the MOD returned by the homogeneous WAA in the sense of preserving its PHD. The first order approximation method has been effectively adopted before in \cite{Vo2009} and \cite{Reuter2014} to achieve the conjugate closure of the MB filter and the labeled MB filter, respectively.
\begin{Rem}
As is indicated in Remark \ref{Rem:HMPHDC14} that the prerequisites C.1--C.4 are all satisfied for the H-MPHD filter. However, it should be noted that, during the derivation of the H-MMB filter, C.1--C.4 need not hold well for all scenarios. For example, if the HPD $\alpha$-credible region of a Bernoulli component is extremely large such that the sensor ability and environmental impact vary significantly within it, then inequality \eqref{eq:HMMB_C2} fails and C.2 does not hold. In these cases, H-MMB filter cannot be applied ideally.
\end{Rem}

\subsection{Evaluation of the Heterogeneous Fusion Weights}\label{sec:HMMB_Weights}
Similarly, the ICCs and heterogeneous fusion weights $\mathcal{W}$ can be calculated using EUF $\Phi(x;e,s)$. For any factorized Bernoulli component $\pi^{\text{\texttt{B}}}_{i,b}$, its ICC can be computed as follows based upon the validation of C.2,
\begin{equation}\label{eq:ICC_HMMB1}
\phi_{i,b}=\left[\Phi_i(x_b^\ast;e_i,s_i)\right]^{-1},\hspace{2mm}x_b^\ast\in S_\alpha(f_{i,b}),
\end{equation}
where $S_\alpha(f_{i,b})$ is the HPD $\alpha$-credible region of $f_{i,b}(x)$ defined in Definition \ref{Def:HPD_Def}, $\Phi_i$ denotes the EUF of the $i$-th sensor, and $x_b^\ast$ can be any state variable in $S_\alpha(f_{i,b})$, e.g., the mode of $f_{i,b}$.
Next, the heterogeneous weights $\mathcal{W}$ can be readily computed by substituting \eqref{eq:ICC_HMMB1} into \eqref{eq:Group FW}.

\subsection{Gaussian Mixture Implementation}
Suppose the GM style MB MOD $\pi^{\text{\texttt{MB}}}_i=\{(r_{i,b},f_{i,b})\}_{b\in \mathbb{B}_{i}}$ is obtained at each local sensor~\cite{Vo2009}, then the spatial distribution $f_{i,b}(x)$ of the $b$-th hypothesized object is represented as
\begin{equation}
\begin{split}\label{eq:GM-MB}
f_{i,b}(x)={\sum}_{j=1}^{J_{i,b}}\alpha_{i,b}^{j}\mathcal{G}\left(x;m_{i,b}^{j},P_{i,b}^{j}\right),
\end{split}
\end{equation}
where ${J_{i,b}}$ and $\alpha_{i,b}^{j}$ denote the number of Gaussian components and the weight of the $j$-th component, respectively. Then, by substituting (\ref{eq:GM-MB}) into (\ref{eq:part fusion Bernoulli2}), the $g$-th, $g=1,\ldots,G_{\mathcal{C}}$, spatial density obtained by the H-MMB filter can be written as,
\begin{align}
\label{eq:GM-MB1}
\bar f_g(x) &=  \sum\limits_{i\in\mathcal{N}, {b\in \mathbb{M}_{i,g}}} \sum_{j=1}^{J_{i,b}} \left(\frac{\omega_{i,g} r_{i,b}\alpha_{i,b}^{j}}{\bar r_g}\right) \mathcal{G}\left(x;m_{i,b}^{j},P_{i,b}^{j}\right).
\end{align}
where ${\bar r_g}$ is the fused existence probability, and can be readily computed using \eqref{eq:part fusion Bernoulli1}. It should be noticed that (\ref{eq:GM-MB1}) remains the structure of Gaussian mixture. In this case, the multi-object state extraction approach of MB filter can be directly applied~\cite{Vo2009,Reuter2014,Vo2010}. Specifically, the number of objects can be estimated by the expected or maximum a posterior cardinality estimate, and the individual states can be extracted by selecting the mean of the corresponding number of Gaussian components with the highest existence probabilities.

After the fusion of \eqref{eq:GM-MB1}, the number of components also increases additively, therefore pruning and merging of components~\cite{Vo2009,Reuter2014,Vo2010} are required to control the component number.

{\subsection{Algorithm Complexity}\label{sec:GM-HMMB-Cost}
Different from the H-MPHD filter, the implementation of the H-MMB filter involves an extra \emph{union find set} based clustering process. The worst-case time complexity of the \emph{union find set} algorithm is about $\mathcal{O}(a\zeta(a,b))$, where $\zeta(a,b)$ denotes a functional inverse of Ackermann's function \cite{Cormen2001}, integers $a$ and $b$ are, respectively, the number of \emph{find} operations and the number of Bernoulli components with $b=\sum_i |\mathbb{B}_i|$. After clustering, we also have to assign the heterogeneous fusion weights to all Bernoulli components, which is with a complexity $\mathcal{O}(b)$. Lastly, the complexity of the weighted averaging for all Gaussian components using \eqref{eq:GM-MB1} is about $c={\sum}_{i\in\mathcal{N}, {b\in \mathbb{M}_{i,g}}}{J_{i,b}}$. Therefore, the total fusion complexity of the H-MMB filter is about $\mathcal{O}(a\zeta(a,b)+b+c)$.}


\section{Performance Assessment}\label{sec:performance}

\begin{figure}[t]
	\centering
	{\includegraphics[width=0.90\columnwidth,draft=false]{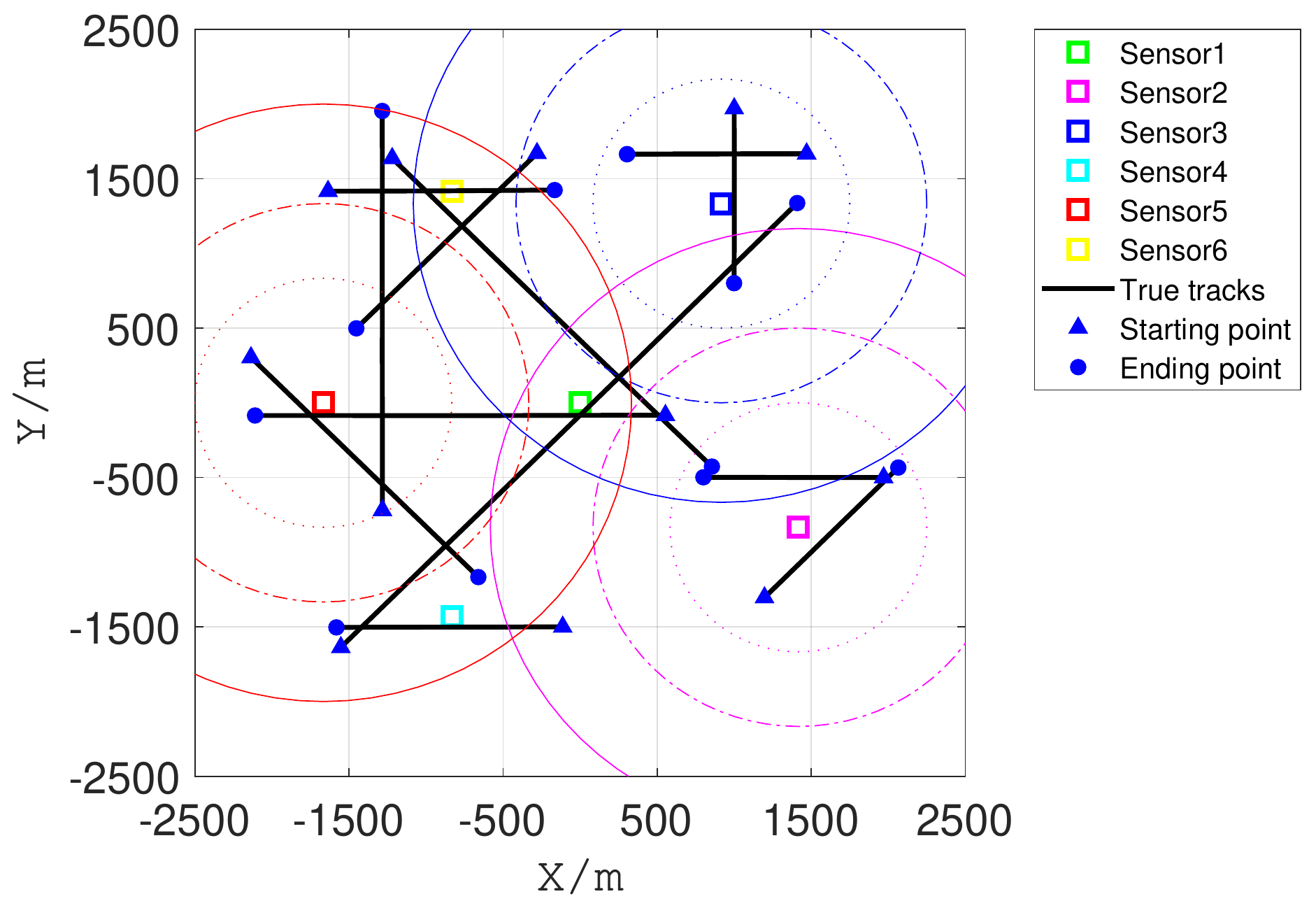}}
	\caption{{The $5000$ m $\times$ $5000$ m 2-dimensional tracking scenario involving $6$ range-angle sensors and $12$ moving objects. The range thresholds for sensors $2$, $3$ and $5$ are illustrated as examples and distinguished with different colors and line types. Specifically, $\Upsilon_{i}^{1}$, $\Upsilon_{i}^{2}$ and $\Upsilon_{i}^{\max}$ are indicated by dotted, dot-segment and solid lines, respectively.}}
	\label{fig:scenario}
\end{figure}

In order to assess the performance of the proposed heterogeneous fusion algorithms, {here we consider tracking $12$ moving objects in a $2$-dimensional surveillance region $\mathbb{S}= \mathbb{L}^2$ with $\mathbb{L}\triangleq[-2500 \text{m},2500 \text{m}]$ using $6$ range-angle radar-type sensors with limited sensing abilities}. The sensor positions and ground truth of all objects are shown in Fig. \ref{fig:scenario}. As for the architecture of the sensor network, at each time, the local MODs of all sensor nodes are sent to a fusion center to compute the global fusion results, and there is no information feedback to each sensor. The performance comparisons with {the state-dependent WAA (SD-WAA) PHD filter~\cite{Yi2020TSPDFoVs}}, and the direct implementation of the homogeneous WAA fusion methods with the Metropolis fusion weights \cite{Xiao2005} are presented to demonstrate the efficacy of the proposed methods. The considered performance metrics include the optimal sub-pattern assignment (OSPA) error \cite{Schumacher}, cardinality statistics {and execution times.} All metrics are averaged over $200$ Monte Carlo runs in the following simulations.

The number of objects in the surveillance area varies over a period of $80$ seconds as new objects can appear and existing objects can disappear. The state of an object is defined as {$x^k=[p_{{\texttt{x}}}^k,p_{\texttt{y}}^k,v_{\texttt{x}}^k,v_{\texttt{y}}^k]^{\top}\in \mathbb{X}$, where the state space $\mathbb{X}\triangleq\mathbb{S} \times  \mathbb{R}^2$, the terms $(p_{\texttt{x}}^k,p_{\texttt{y}}^k)\in \mathbb{S}$ and $(v_{\texttt{x}}^k,v_{\texttt{y}}^k)\in\mathbb{R}^2$ represent the position and velocity in the ${\texttt{x}}$ and ${\texttt{y}}$ coordinates}, respectively. The single-object dynamic model $\varphi^{k|k-1}(\cdot | \cdot )$ is assumed to be linear and Gaussian as,
\begin{equation}
\varphi^{k|k-1}(x^{k}|x^{k-1}) = {\cal G}(x^{k};Fx^{k-1},Q),
\end{equation}
with
\begin{equation}
\begin{aligned}
{F} = \left[ {\begin{array}{*{20}{c}}
	{{1}}\\
	{{0}}
	\end{array}\begin{array}{*{20}{c}}
	{}\\
	\end{array}\begin{array}{*{20}{c}}
	{ {T_s}}\\
	{{1}}
	\end{array}} \right]\otimes {I_2},
\quad {Q} = \sigma _v^2\left[ {\begin{array}{*{20}{c}}
	{{\textstyle{{{T_s ^3}} \over 3}}} \quad {{\textstyle{{{T_s ^2}} \over 2}}}\\
	{{\textstyle{{{T_s ^2}} \over 2}}} \quad {{T_s }}
	\end{array}} \right]\otimes {I_2},
\end{aligned}
\end{equation}
where the sampling interval $T_s=1$~s, the standard deviation of the process noise $\sigma_v =0.1$ m/s$^2$, $I_n$ is an $n\times n$ identity matrix, and $\otimes$ denotes the Kronecker product operation. Moreover, the probability of object survival is set as $p_S=0.98$.

Each sensor $i\in\{1,2,3,4,5,6\}$ has a circle-shaped FoV with maximum detection range $\Upsilon_{i}^{\max}$ due to its physical limitation of sensing ability. For all sensors, the number of clutter measurements at each sampling time is Poisson distributed with parameter $\lambda_c = 5$, and each clutter is spread uniformly over the sensor FoV. The $i$-th sensor can only collect the valid measurements of a object when the object is inside its FoV,
\begin{equation}\label{eq:FoV_func}
\digamma_i({x}^k)  = \left\{ \begin{array}{l}
1, \hspace{3mm}  d(s_i,x^k)\le\Upsilon_{i}^{\max} \\
0, \hspace{3mm}  d(s_i,x^k)>\Upsilon_{i}^{\max},
\end{array} \right.
\end{equation}
where $d(s_i,x^k)=\sqrt {{{({p_{\texttt{x}}^k} - p_{i,{\texttt{x}}})}^2} + {{({p_{\texttt{y}}^k} - p_{i,{\texttt{y}}})}^2}}$ with $s_i=[p_{i,{\texttt{x}}},p_{i,{\texttt{y}}}]^{\top}$ the location of the $i$-th sensor, and recall $\digamma_i(\cdot)$ is the FoV function of sensor $i$ given in \eqref{eq:CEUF1}. On the other hand, when object $x^k$ is inside the FoV, sensor $i$ can detect it with a state-dependent probability function as follows,
{
\begin{equation}\label{eq:PD func}
p_{D,i}(x^k) = \left\{ {\begin{array}{*{20}{l}}
	0.98,\hspace{6.8mm}  0\le d(s_i,x^k)\le \Upsilon_{i}^{1}   \\
	0.8,\hspace{5.9mm}   \Upsilon_{i}^{1}\le d(s_i,x^k)\le \Upsilon_{i}^{2}  \\
	0.6, \hspace{5.9mm}   \Upsilon_{i}^{2}\le d(s_i,x^k)\le \Upsilon_{i}^{\max},
	\end{array}} \right.
\end{equation}
where $\Upsilon_{i}^{1}=500$, $\Upsilon_{i}^{2}=800$ and $\Upsilon_{i}^{\max}=1200$ are the range thresholds}, and they reflect the different sensing abilities of the local sensors. Each target-originated measurement is generated according to a nonlinear measurement model with zero-mean additive Gaussian measurement noise,
\begin{equation}\label{eq:NonlinearMM}
{z_i^k} = \left[ {\begin{array}{*{20}{c}}
	{\text{atan2}({p_{\texttt{x}}^k} - p_{i,{\texttt{x}}},{p_{\texttt{y}}^k} - p_{i,{\texttt{y}}})}\\
	{d(s_i,x^k) }
	\end{array}} \right] + {\nu ^k},
\end{equation}
where atan2$(\cdot)$ denotes the $4$-quadrant inverse tangent function and the covariance matrix of the measurement noise $\nu^k$ is
\begin{equation}\label{eq:SensorMN}
R^k = \left[
        \begin{array}{cc}
          \sigma _\theta ^2 & 0 \\
          0 & \sigma_r^2 \\
        \end{array}
      \right]
\end{equation}
with $\sigma_{\theta}=2^\circ$ rad and {$\sigma_{r}=20$m}. Due to the nonlinear measurement model \eqref{eq:NonlinearMM}, the extended Kalman version of the PHD~\cite{Vo2006} and MB~\cite{Vo2009} filters are employed in the simulation.

\subsection{Case 1}
Before fusion, the normalized heterogeneous weights need to be computed for the H-MPHD and H-MMB filters. {As for the implementation of the H-MPHD filter, we need to begin with obtaining an $M$-portion space partition $\mathcal{X}_M(\mathbb{X})$, according to Definition \ref{Def:Space-partition}, such that C.2 can be satisfied. Since the sensor abilities only vary along the position dimensions in this scenario, we consider the space partition only in the position dimensions. Specifically, the $2$-dimensional surveillance region $\mathbb{S}$ is evenly partitioned into $M=2500^2$ disjoint square-shape sub-regions $\{\mathbb{S}_m\}_{m=1}^M$ with a same $2\text{m}\times2\text{m}$ size. Namely, $\mathbb{S}=\cup_{m=1}^M  \mathbb{S}_m$, and $\mathbb{S}_n\cap \mathbb{S}_m=\emptyset, \hspace{2mm}n,m \in \{1, \ldots, M\},\hspace{1mm} n \neq m$. Based on the partition of $\mathbb{S}$, the adopted partition of $\mathbb{X}\triangleq\mathbb{S} \times  \mathbb{R}^2$ can be readily expressed as $\mathbb{X}=\cup_{m=1}^M \mathbb{X}_m$ with the $m$-th subspace $\mathbb{X}_m=\mathbb{S}_m\times  \mathbb{R}^2$. Given the sensor characteristics described by \eqref{eq:SensorMN}, C.2 can be satisfied for such a $2\text{m}\times2\text{m}$ size partition. Namely, the information confidence within each sub-MOD $\pi_{i,m}$ is relatively consistent and can be quantified by a scalar-type ICC $\phi_{i,m}$.}

{Let ${x_m^\ast}$, $m=1,\ldots,M$, denote the central point of the $m$-th subspace $\mathbb{X}_m$. The CEUF
$\Phi^c(x_m^{\ast},e_i,s_i)$ can be computed for each $\pi_{i,m}$ by substituting \eqref{eq:FoV_func} and \eqref{eq:PD func} into \eqref{eq:CEUF1}. Next, the AEUF $\Phi^a(x_m^{\ast},e_i,s_i)$ is evaluated as $|R_m^x(z_m^\ast)|_\text{d}$ where $z_m^\ast$ is the ideal measurement generated by ${x_m^\ast}$ using (\ref{eq:NonlinearMM}) when the measurement noise is set as zero.
The term $R_m^x(z_m^\ast)$ denotes the position related coordinate-converted measurement covariance matrix of $z_m^\ast$~\cite{lerro1993tracking}. Thus, $R_m^x(z_m^\ast)$ is a matrix of size $2 \times 2$ and can be computed offline when ${x_m^\ast}$ is chosen.
By substituting the resultant $\Phi^a$ and $\Phi^c$ into \eqref{eq:EUF}, \eqref{eq:ICC_PHD1} and \eqref{eq:omega_PHD1}, the heterogeneous weights $\{\omega_i(x_m^{\ast})\}_{i=1}^6$ of \eqref{eq:H-MS-PHD_omega} can be obtained for the H-MPHD filter. The coefficients in Definition $4$ are set as $u_1=1$ m$^{-4}$ and $u_2=800$ to balance the relative penalties of AEUF and CEUF since $|R_m^x(z_m^\ast)|_\text{d}$ takes much larger value than detection probability.}
\begin{figure}[t]
	\centering
	\subfigure[]{\includegraphics[width=0.44\columnwidth,draft=false]{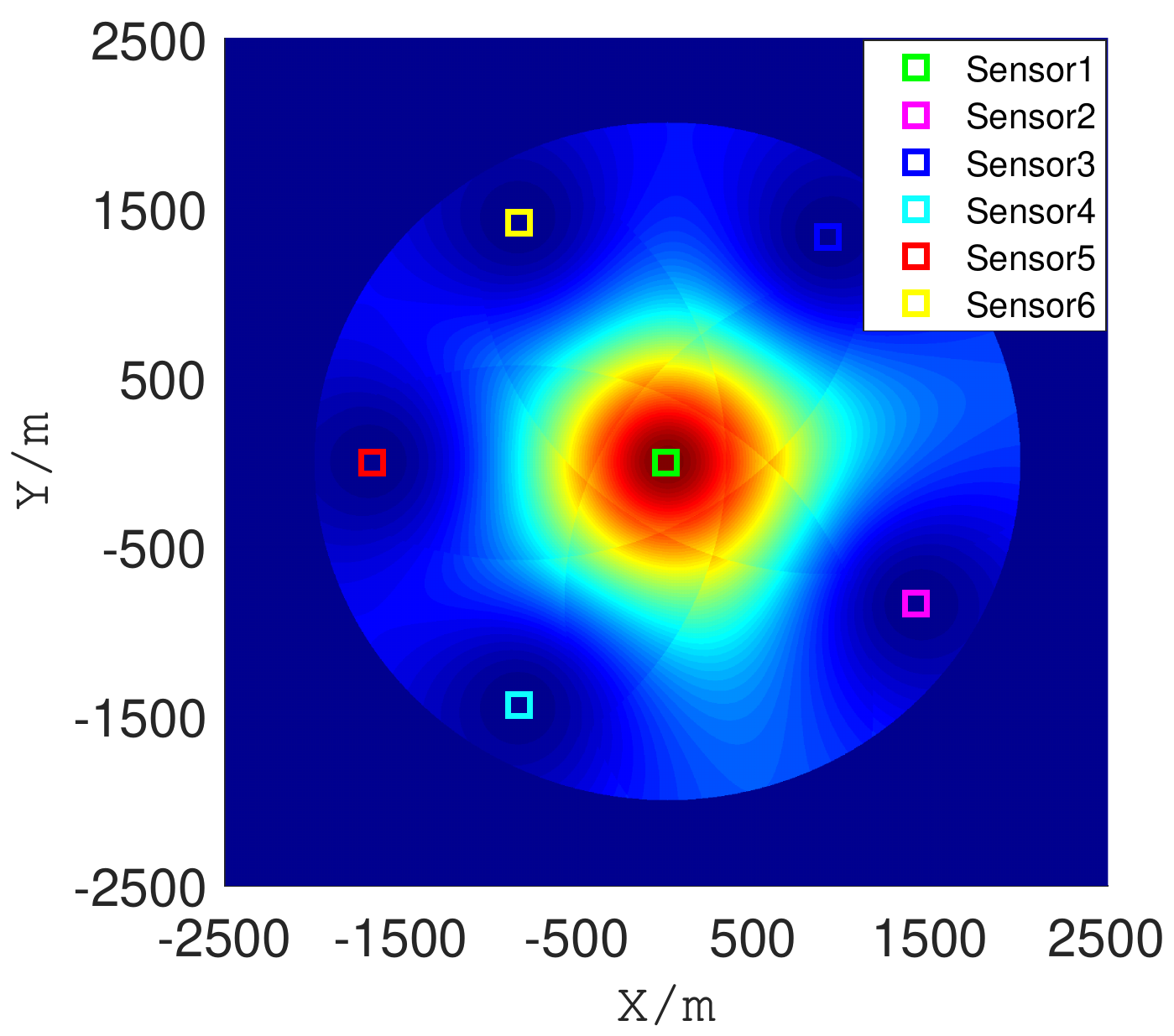}}
	\subfigure[]{\includegraphics[width=0.49\columnwidth,draft=false]{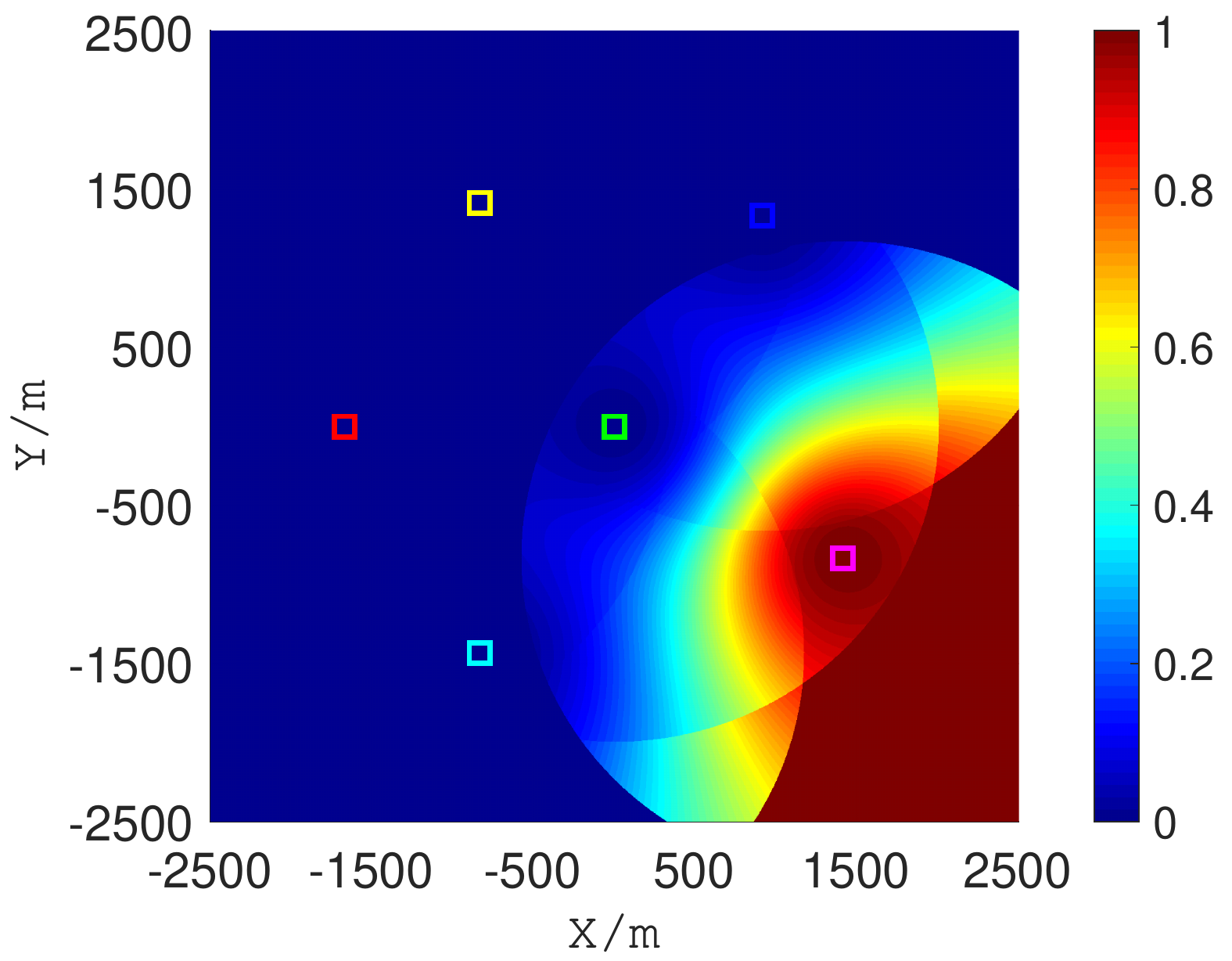}}
	\subfigure[]{\includegraphics[width=0.44\columnwidth,draft=false]{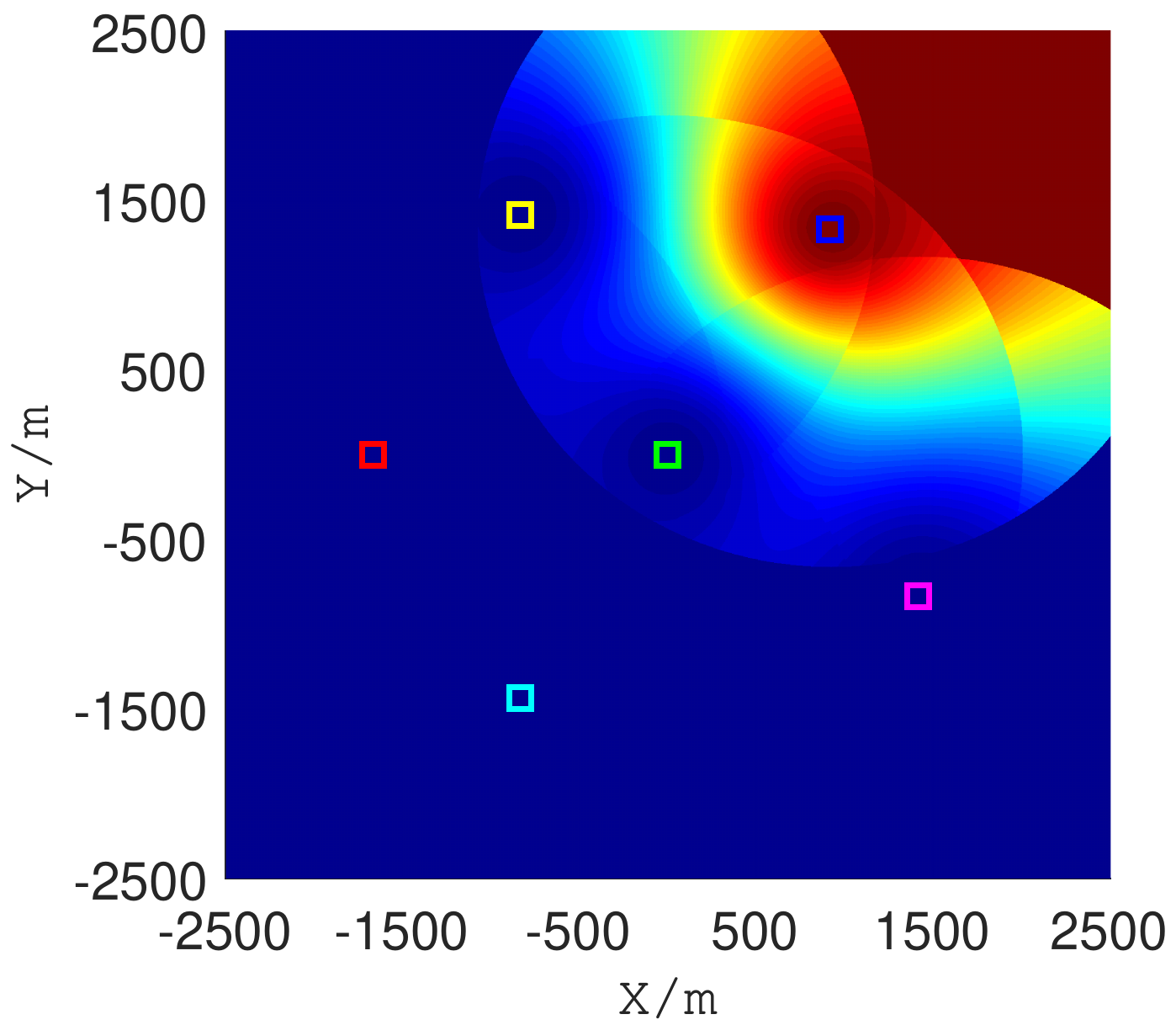}}
	\subfigure[]{\includegraphics[width=0.49\columnwidth,draft=false]{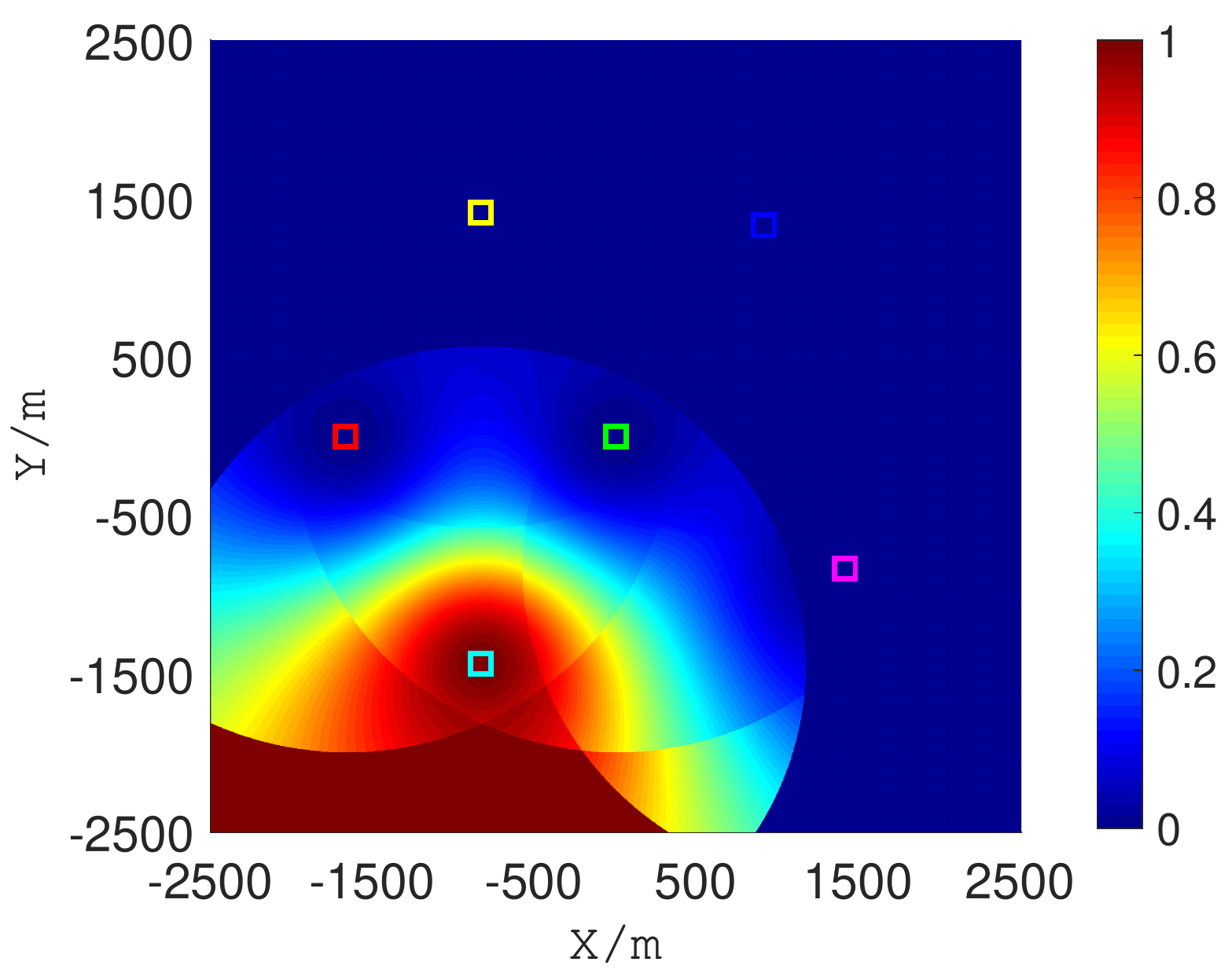}}
	\subfigure[]{\includegraphics[width=0.44\columnwidth,draft=false]{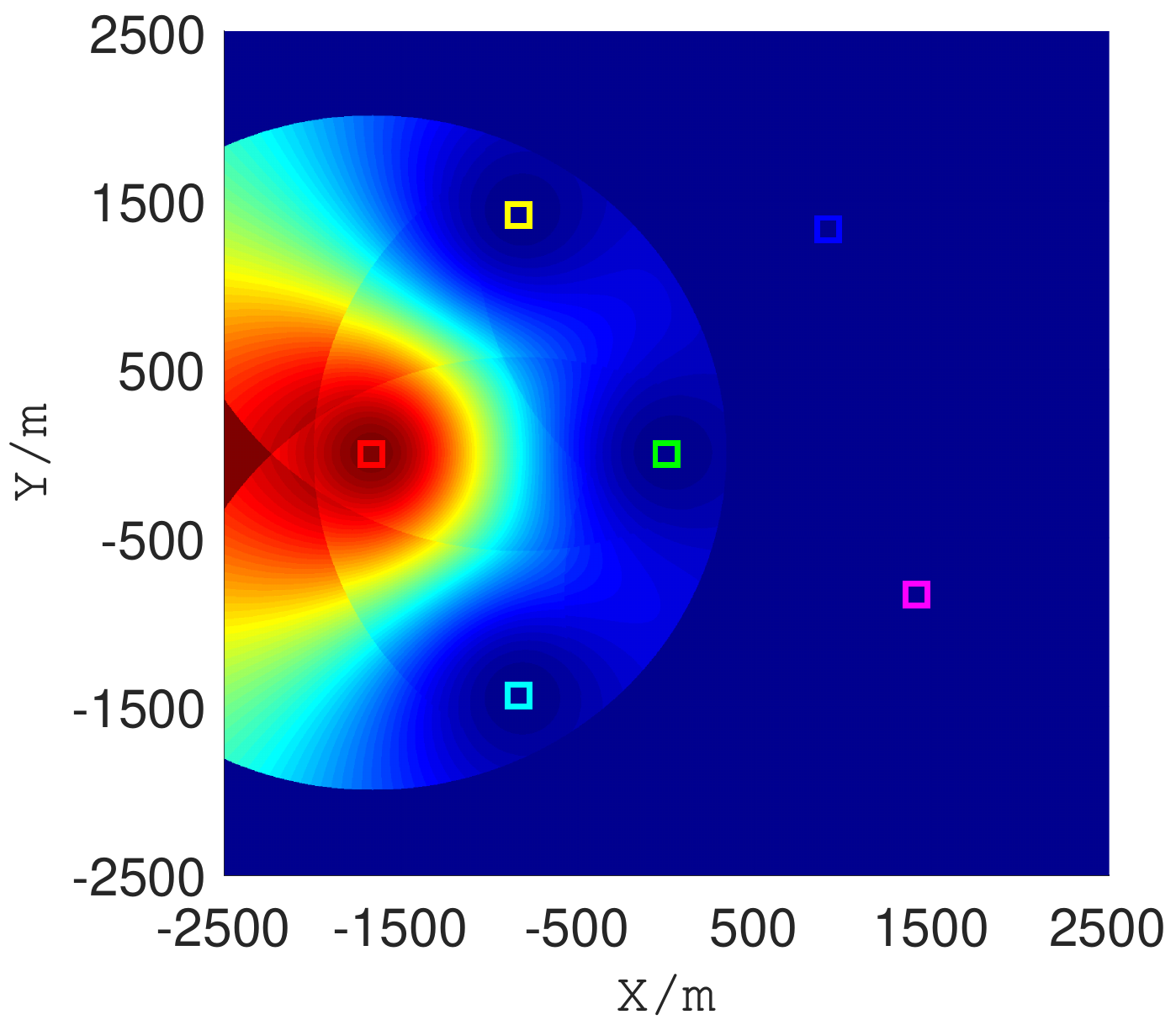}}
	\subfigure[]{\includegraphics[width=0.49\columnwidth,draft=false]{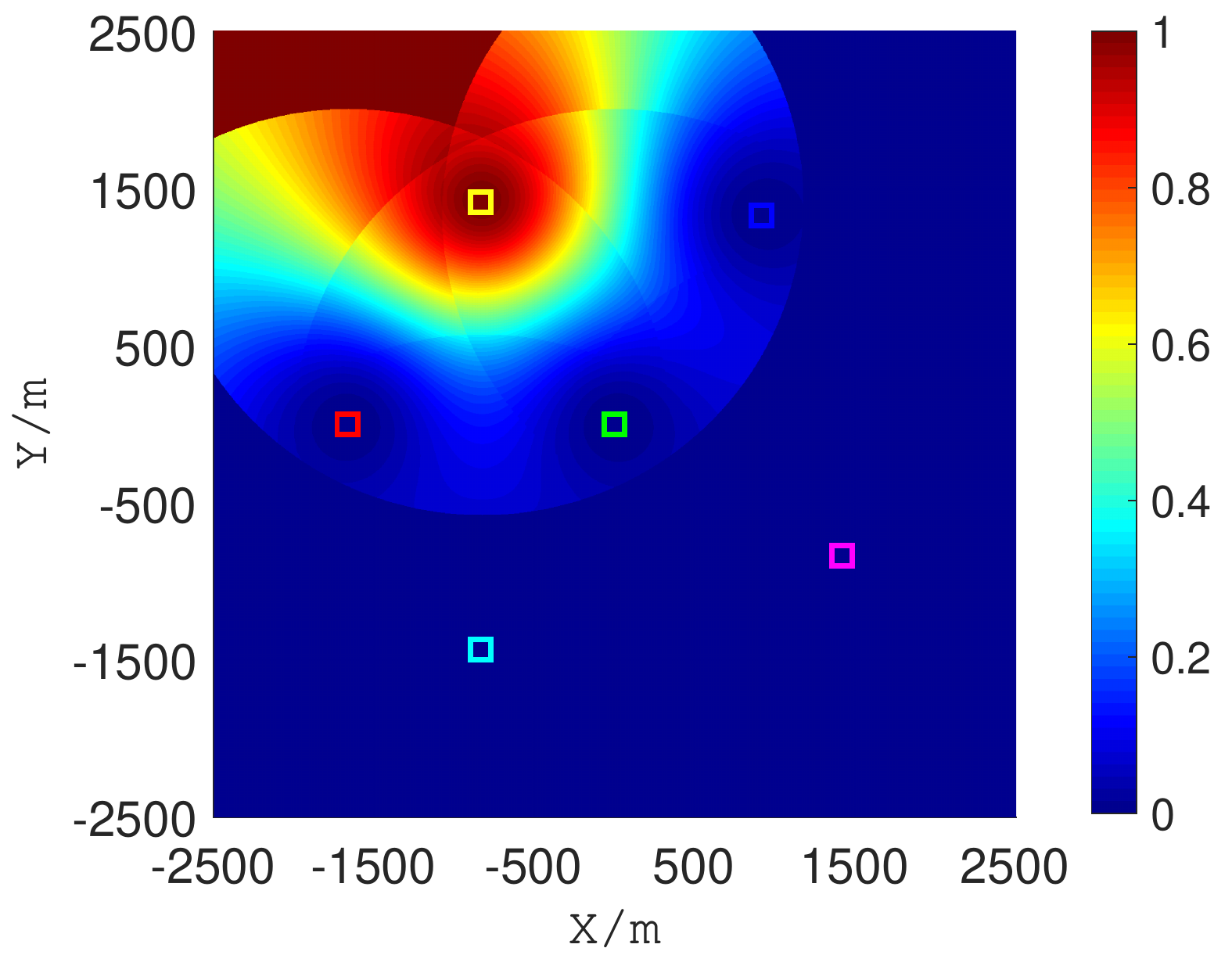}}
	\caption{{Case 1: the intensity map of the heterogeneous fusion weights of all the local sensors are plotted against the ${\texttt{x}}$ and ${\texttt{y}}$ coordinates: (a) Sensor 1, (b) Sensor 2, (c) Sensor 3, (d) Sensor 4, (e) Sensor 5, (f) Sensor 6.}}
	\label{fig:weight}
\end{figure}

{Fig. \ref{fig:weight} shows the heterogeneous fusion weights $\{\omega_i(x_m^{\ast})\}_{i=1}^6$ for the $M=2500^2$ groups of the decomposed sub-MODs $\{\{\pi_{i,m}\}_{i=1}^6\}_{m=1}^M$. Note that the sub-MODs are grouped automatically without using any extra clustering process since all the local MODs are factorized based on a same space partition $\mathcal{X}_M(\mathbb{X})$.}
The values of these weights are indicated by different colors. It can be seen that, in order to reflect the complex relative information confidence among the MODs generated by different sensors, the heterogeneous weights of each sensor are also space-varying. Specifically, for any state $x_m^{\ast}$, the $i$-th sensor has a larger heterogeneous weight $\omega_i(x_m^{\ast})$ when $d(s_i,x_m^{\ast})$ is small. This is because detection probability function $p_{D,i}(x_m^{\ast})$ is inversely proportional to $d(s_i,x_m^{\ast})$, and the value of $|R_m^x(z_m^\ast)|_\text{d}$ increases as $d(s_i,x_m^{\ast})$ increases due to the fixed standard deviation of angular measurement. Also, the value of the heterogeneous weights of a sensor are affected by all the rest sensors. The weight $\omega_i(x_m^{\ast})$ of sensor $i$ is small when other sensors have high estimation confidence on state $x_m^{\ast}$. Particularly, sensor $i$ has full discourse on $x_m^{\ast}$, i.e., $\omega_i(x_m^{\ast})=1$, if $x_m^{\ast}$ is out the FoVs of all other sensors. The results in Fig. \ref{fig:weight} is also intuitive since it is natural to trust a sensor more if the object is close to that sensor, and vice versa. In contrast, the homogeneous WAA method employs the scalar weights without regard for the practical characteristics of sensors.

{As for the implementation of the H-MMB filter, the corresponding heterogeneous fusion weights can also be obtained using $\{\omega_i(x_m^{\ast})\}_{i=1}^6$. In particular, here we select the suitable $\omega_i(x_m^{\ast})$, which is located in the the HPD $\alpha$-credible region $S_\alpha(f_{i,b})$ of Bernoulli density $f_{i,b}(x)$ as the heterogeneous fusion weight for $f_{i,b}(x)$.} {More importantly, it is worth noting that the heterogeneous fusion weights $\{\omega_i(x_m^{\ast})\}_{i=1}^6$ can be computed offline and stored as knowledge at the fusion center. In this case, the fusion weights do not need to be repeatedly computed and transmitted among sensors at each fusion interval for both the H-MPHD and H-MMB filters.}

\begin{figure}[t]
	\centering
	\subfigure[]{\includegraphics[width=0.49\columnwidth,draft=false]{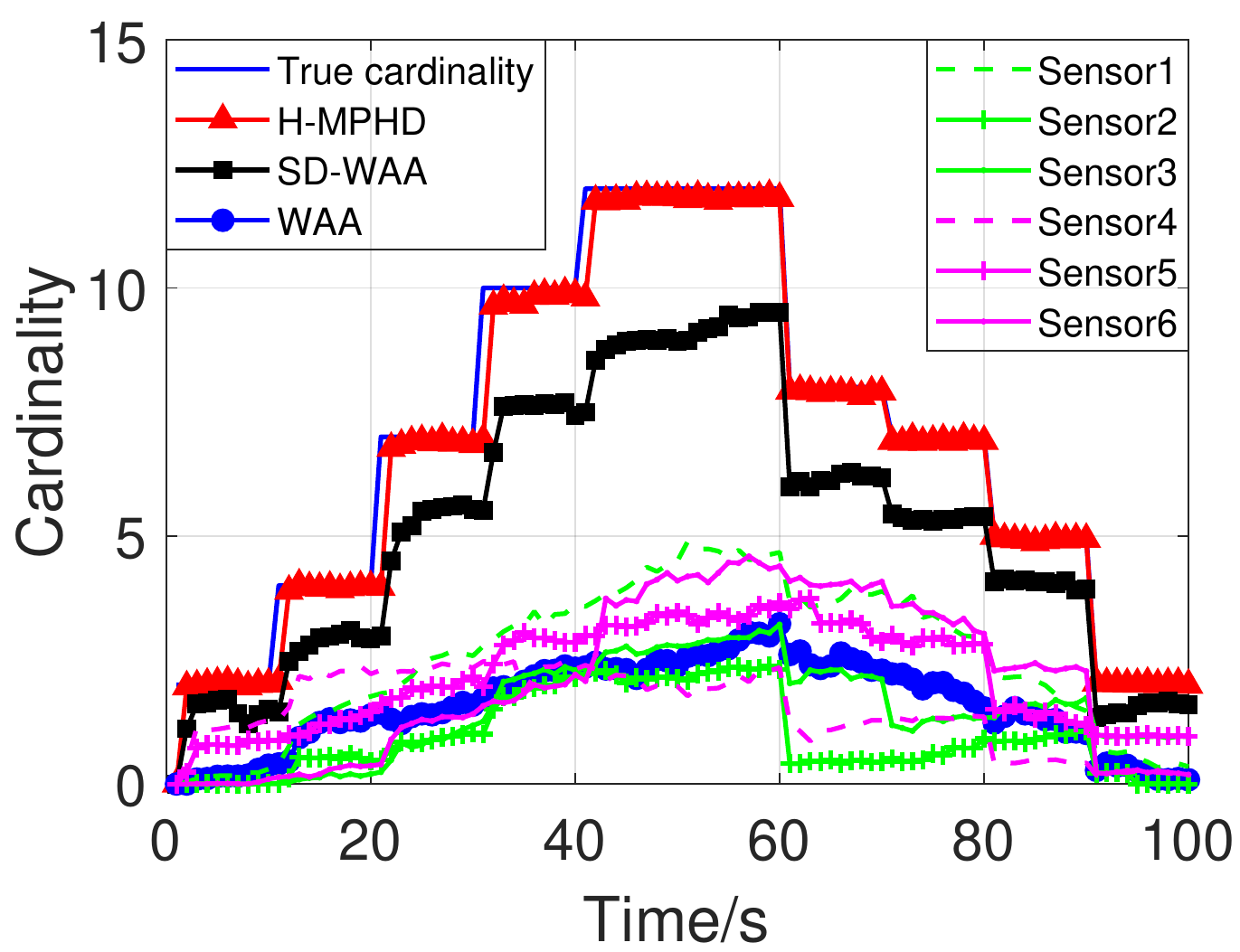}}
	\subfigure[]{\includegraphics[width=0.49\columnwidth,draft=false]{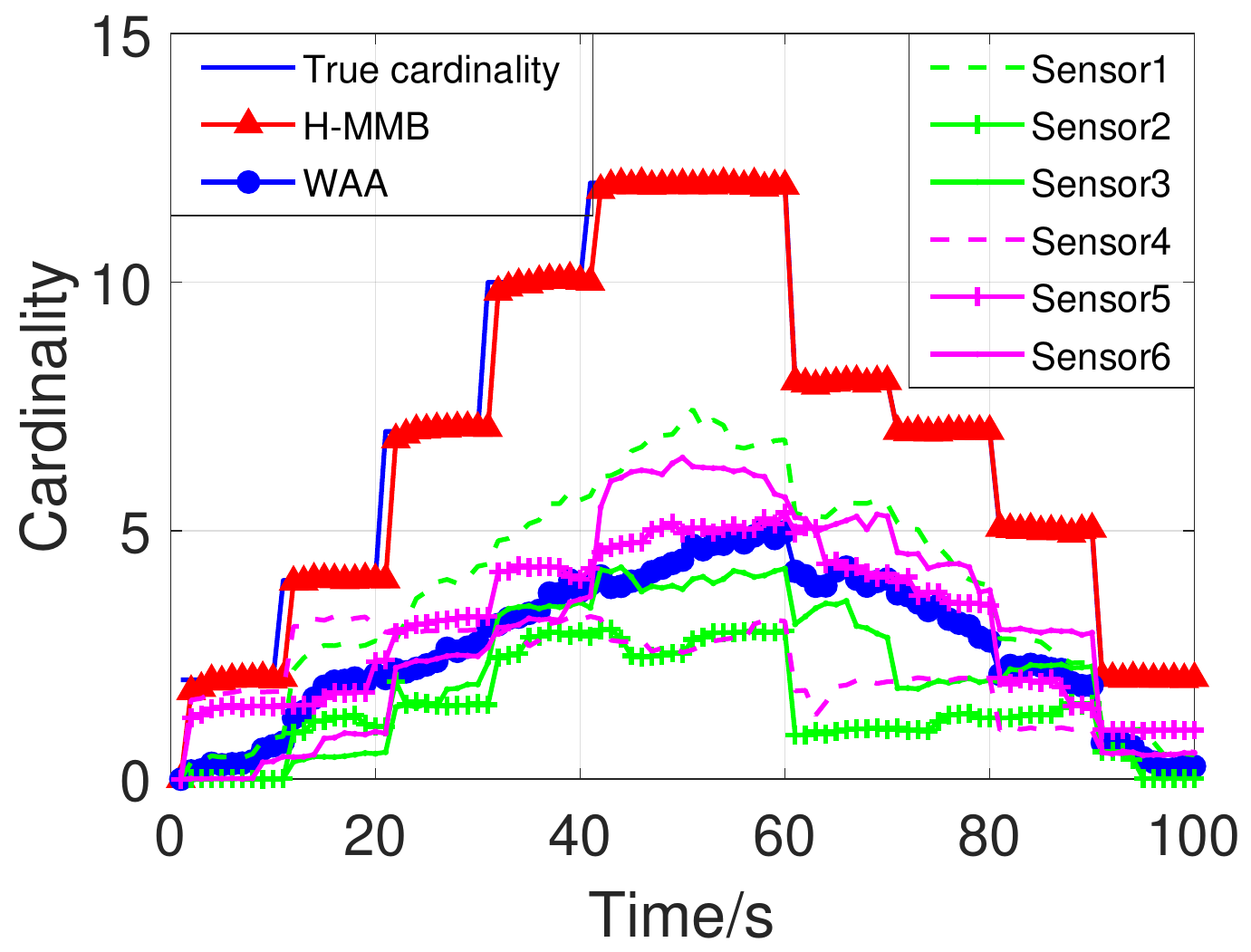}}
	\caption{{Case 1 : the cardinalities are plotted against time for all the single and multi-sensor PHD and MB filters, (a) the PHD filters, (b) the MB filters.}}
	\label{fig:cardi_case1}
\end{figure}

\begin{figure}[t]
	\centering
	\subfigure[]{\includegraphics[width=0.49\columnwidth,draft=false]{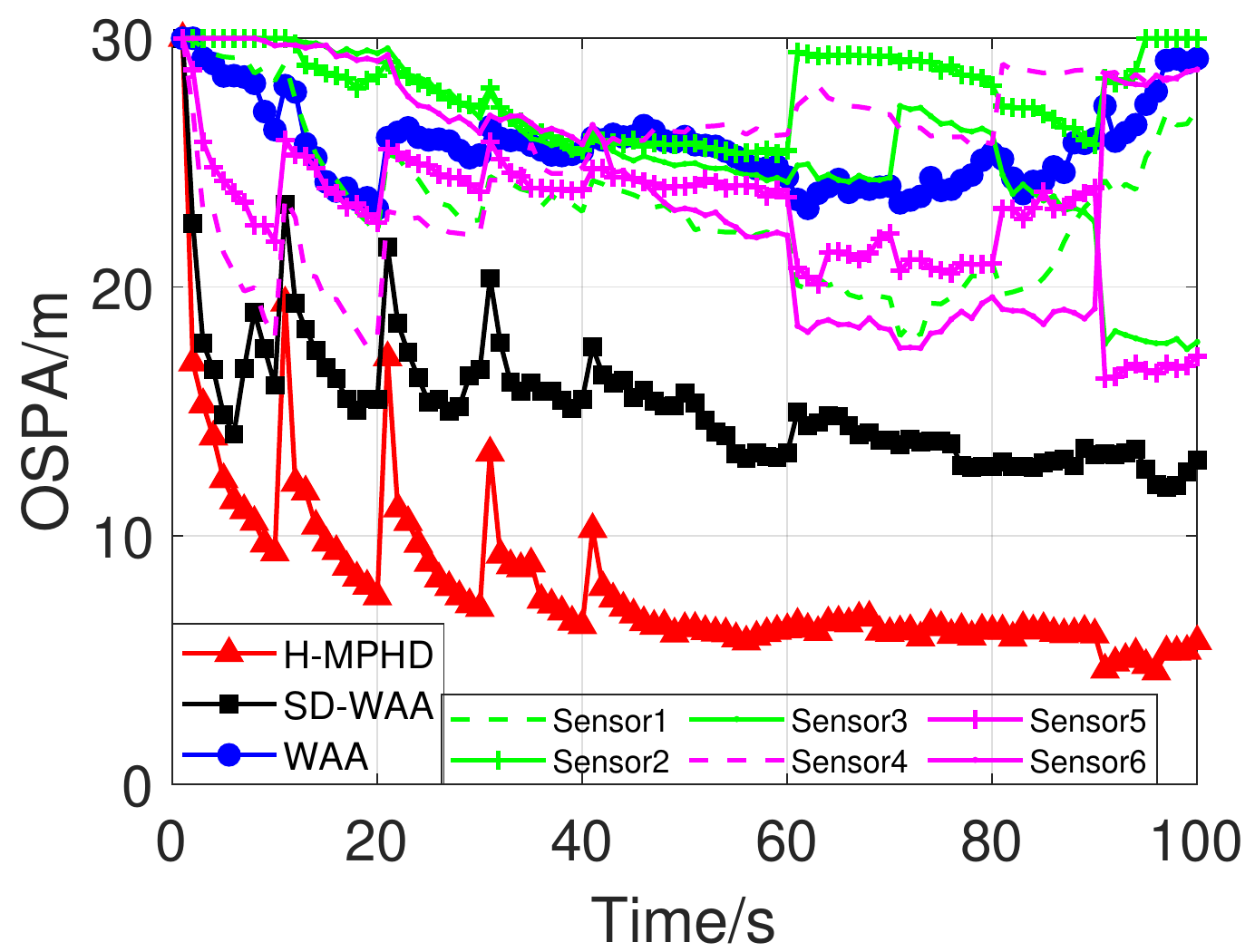}}
	\subfigure[]{\includegraphics[width=0.49\columnwidth,draft=false]{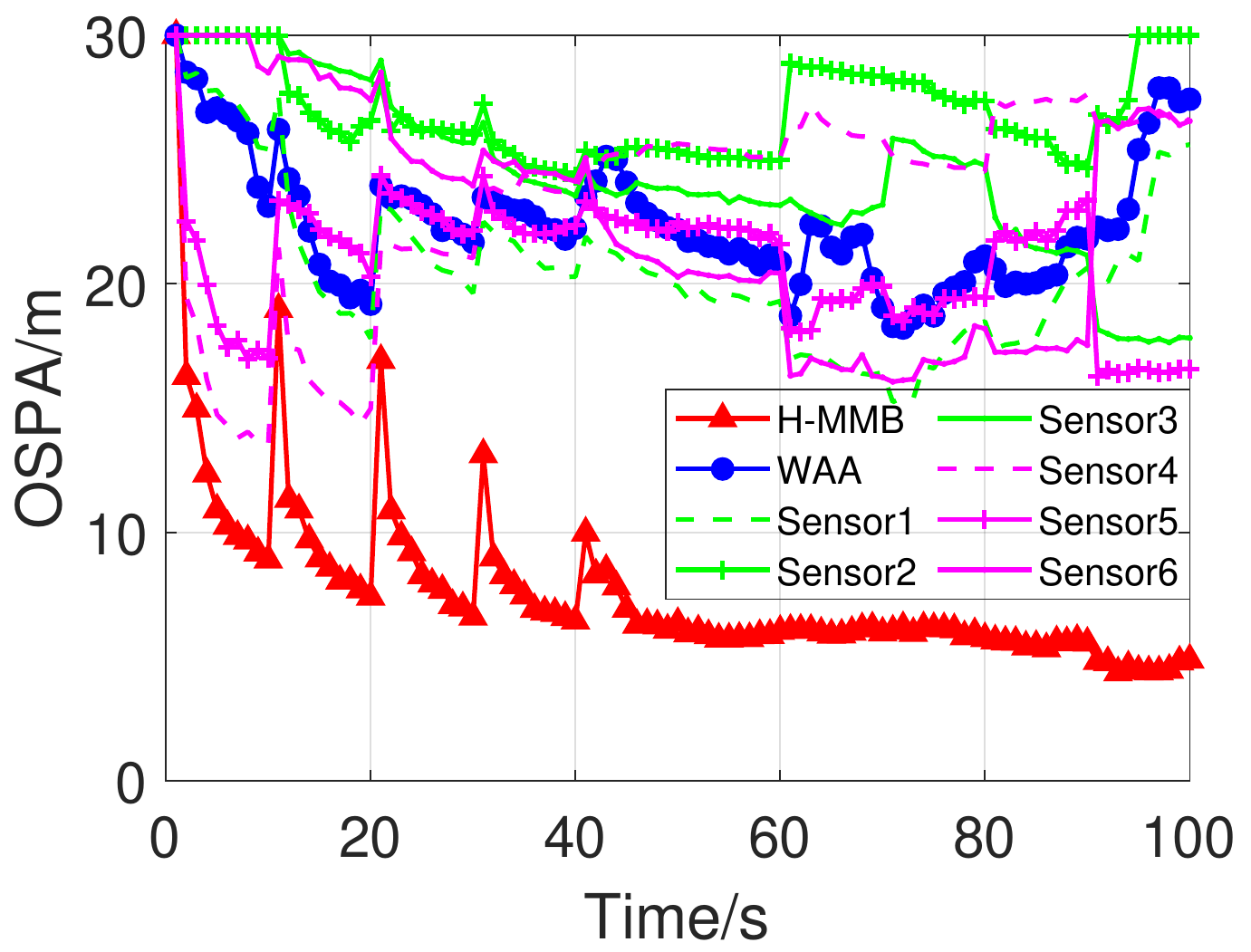}}
	\caption{{Case 1 : the OSPA errors are plotted against time for all the single and multi-sensor PHD and MB filters, (a) the PHD filters, (b) the MB filters.}}
	\label{fig:ospa_case1}
\end{figure}

In Figs. \ref{fig:cardi_case1} and \ref{fig:ospa_case1}, the cardinality and OSPA errors are plotted against time steps for each single sensor, {the SD-WAA PHD filter~\cite{Yi2020TSPDFoVs}}, the H-MPHD and H-MMB filters, and their homogeneous versions.
It can be seen that the single sensors perform poorly due to their limited sensing abilities and FoVs. The standard WAA fusion methods return better results than the local filters due to the combination and complementation of the information collected within the FoVs of individual sensors. {The SD-WAA PHD filter is able to further improve the fusion performance by properly using the knowledge of sensor FoV during the information averaging.} However, they are still incapable of tracking all the objects effectively due to their oversimplified weighting mechanism which fails to reflect relative confidence of MODs in this scenario. In contrast, the proposed H-MPHD and H-MMB filters can effectively track all objects and deliver the best fusion performance.

\subsection{Case 2}
\begin{figure}[t]
	\centering
	\subfigure[]{\includegraphics[width=0.445\columnwidth,draft=false]{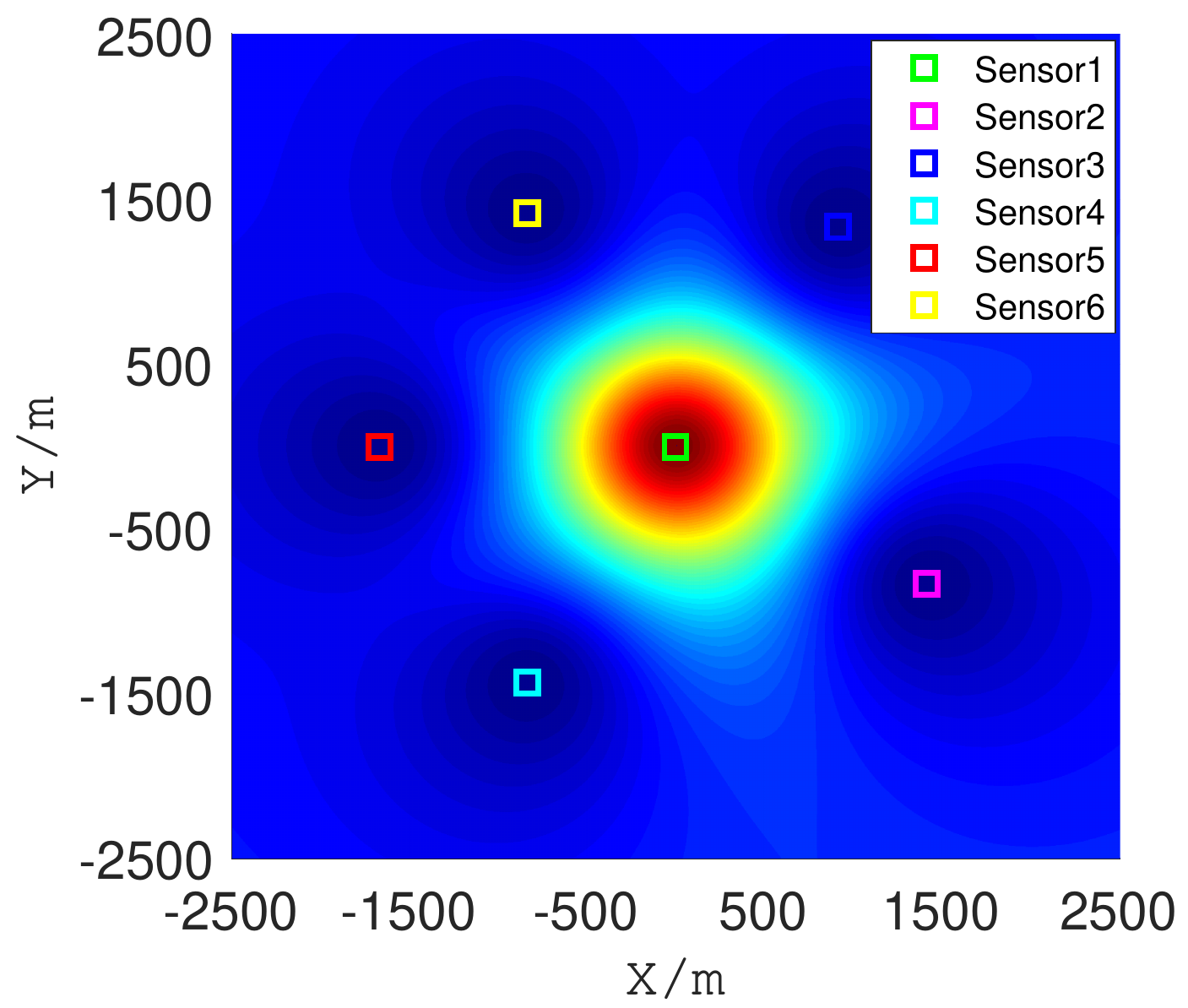}}
	\subfigure[]{\includegraphics[width=0.49\columnwidth,draft=false]{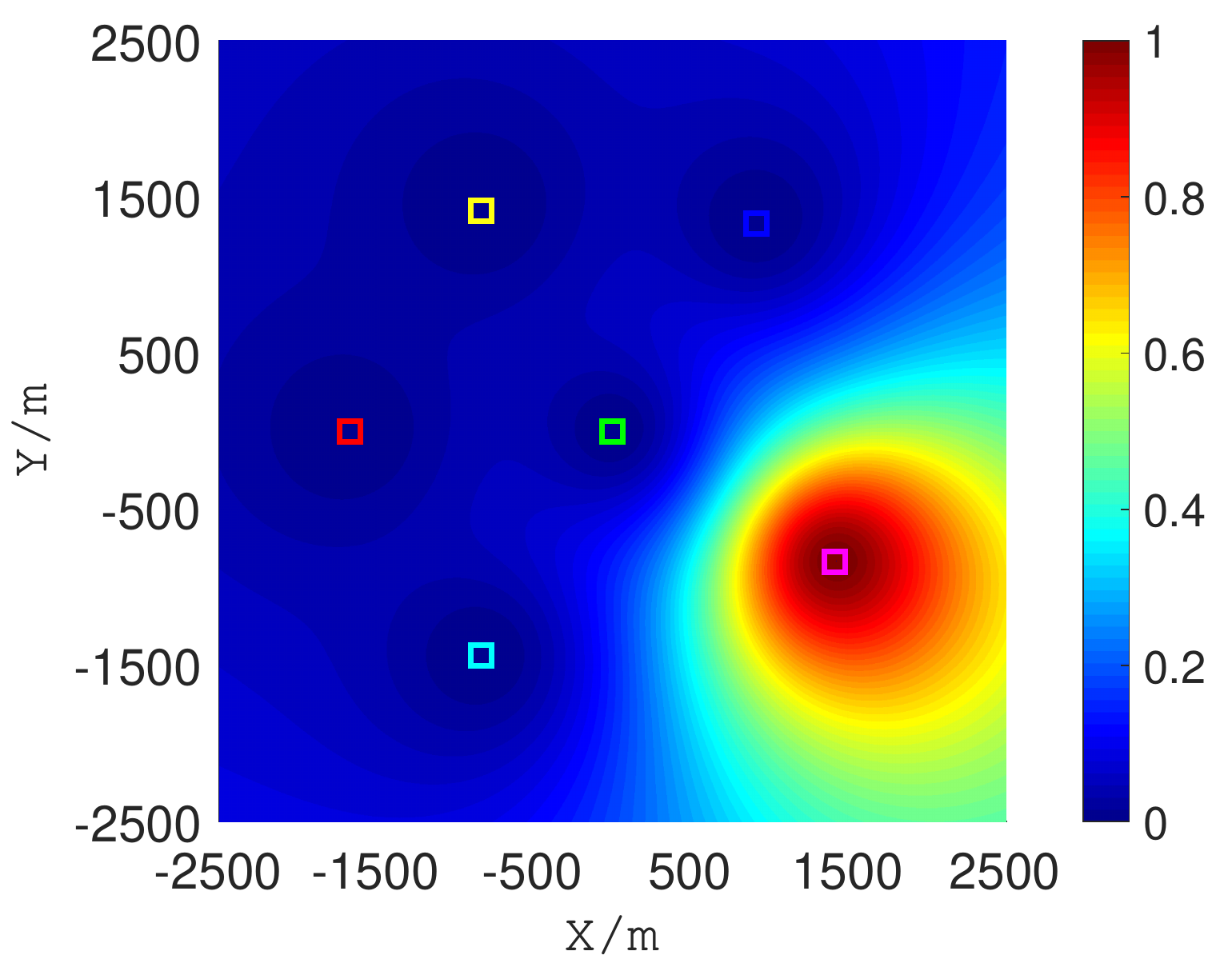}}
	\subfigure[]{\includegraphics[width=0.445\columnwidth,draft=false]{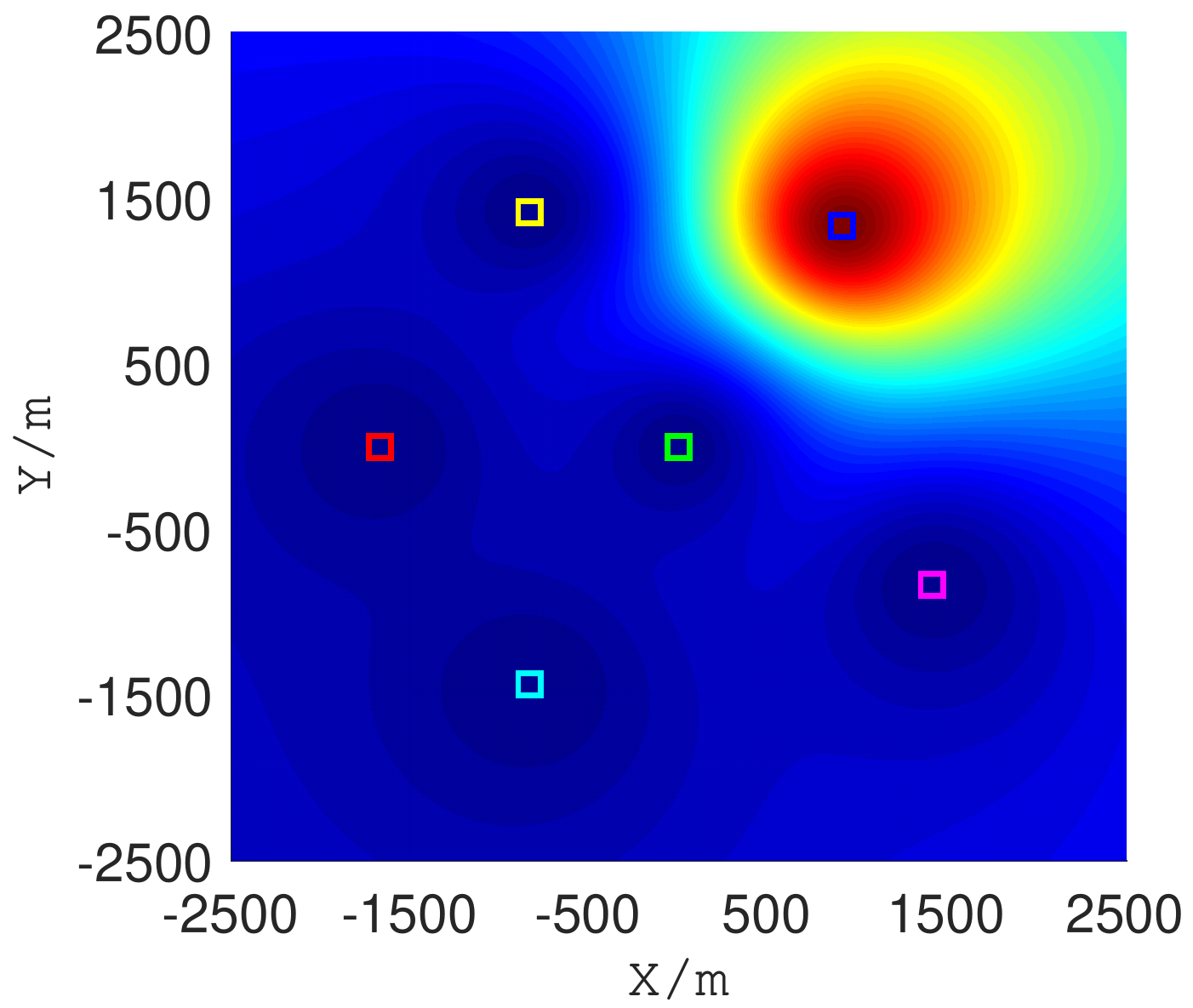}}
	\subfigure[]{\includegraphics[width=0.49\columnwidth,draft=false]{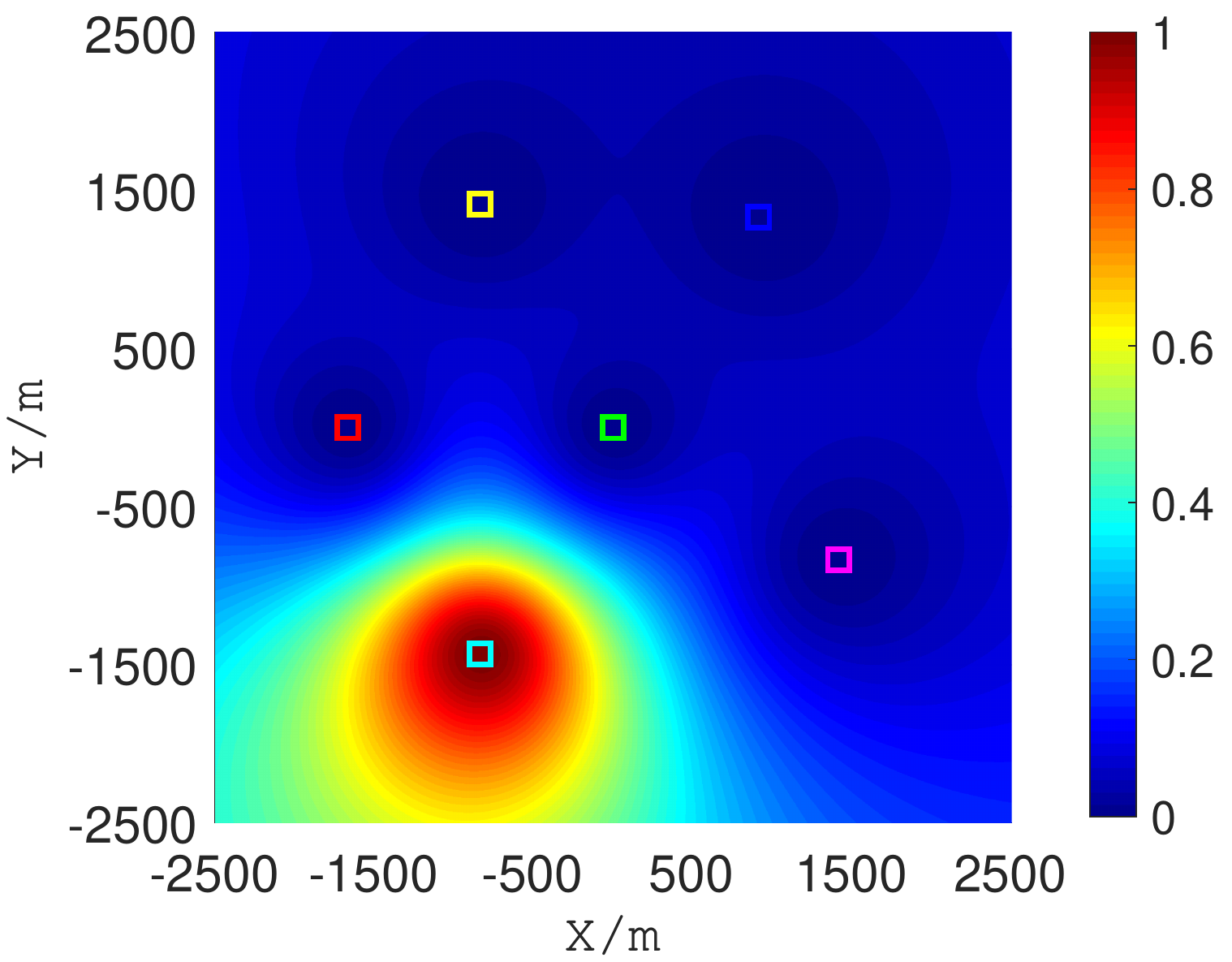}}
	\subfigure[]{\includegraphics[width=0.445\columnwidth,draft=false]{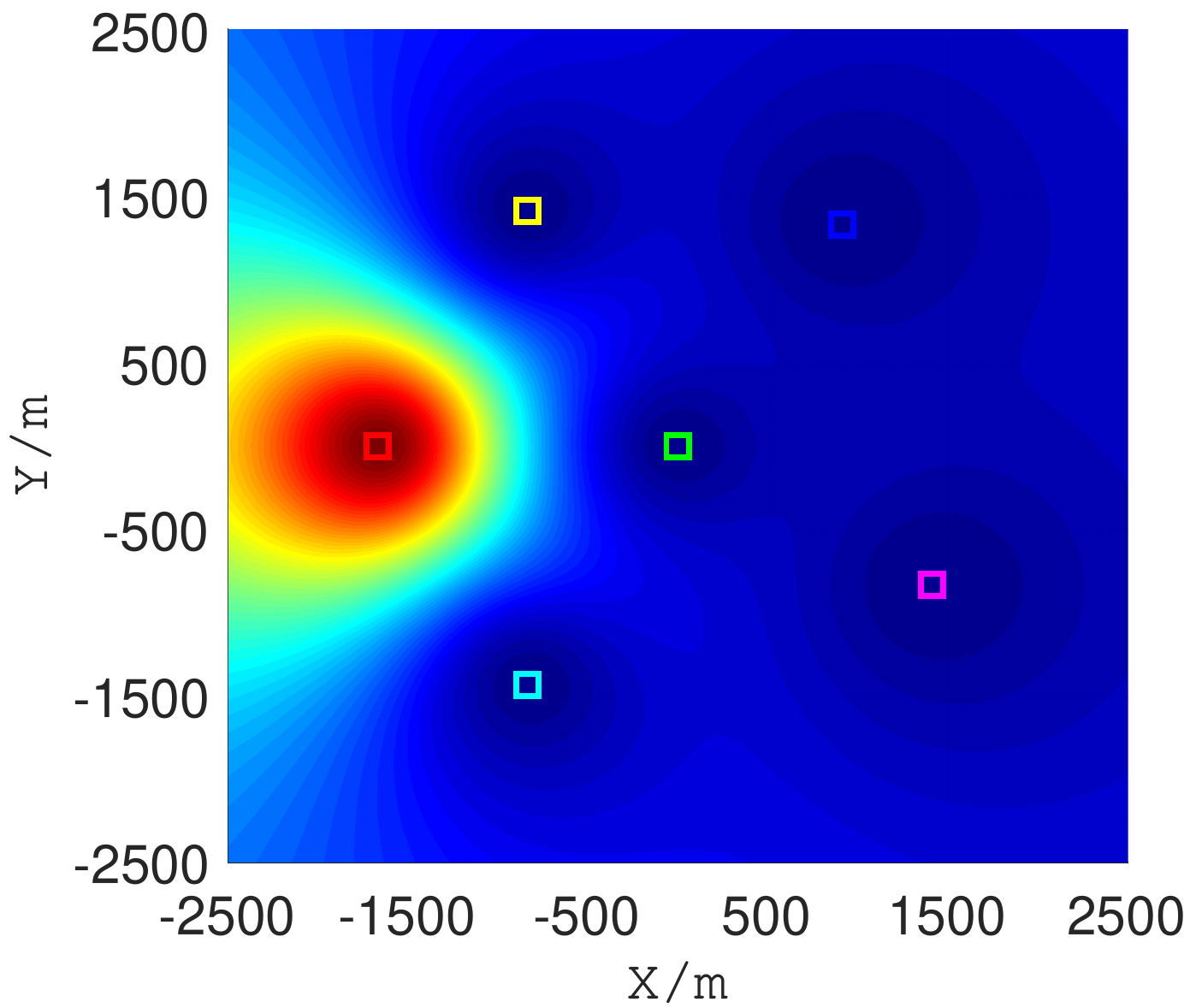}}
	\subfigure[]{\includegraphics[width=0.49\columnwidth,draft=false]{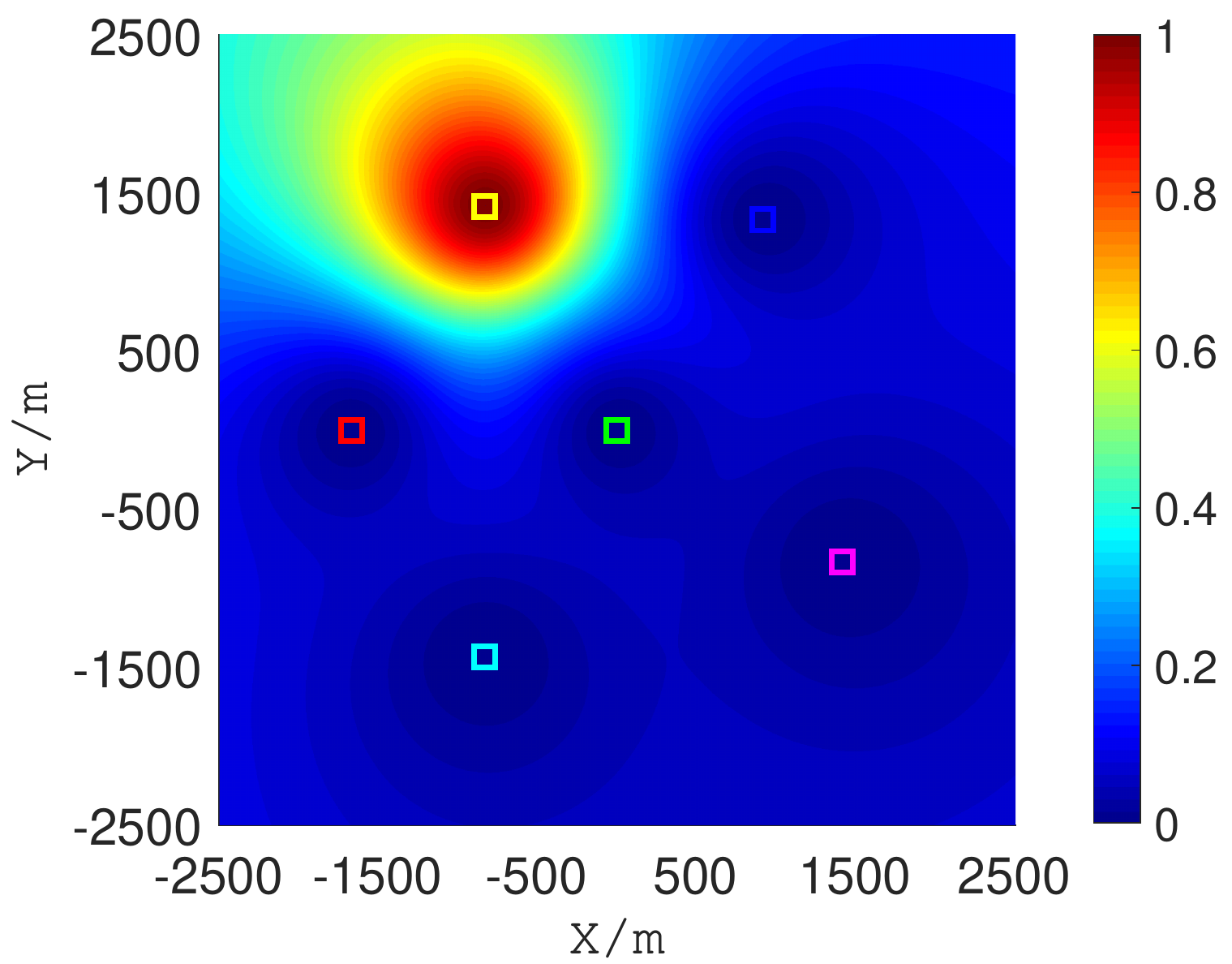}}
	\caption{{Case 2: the intensity map of the heterogeneous fusion weights of all the local sensors are plotted against the ${\texttt{x}}$ and ${\texttt{y}}$ coordinates: (a) Sensor 1, (b) Sensor 2, (c) Sensor 3, (d) Sensor 4, (e) Sensor 5, (f) Sensor 6.}}
	\label{fig:weight-acc}
\end{figure}
\begin{figure}[t]
	\centering
	\subfigure[]{\includegraphics[width=0.49\columnwidth,draft=false]{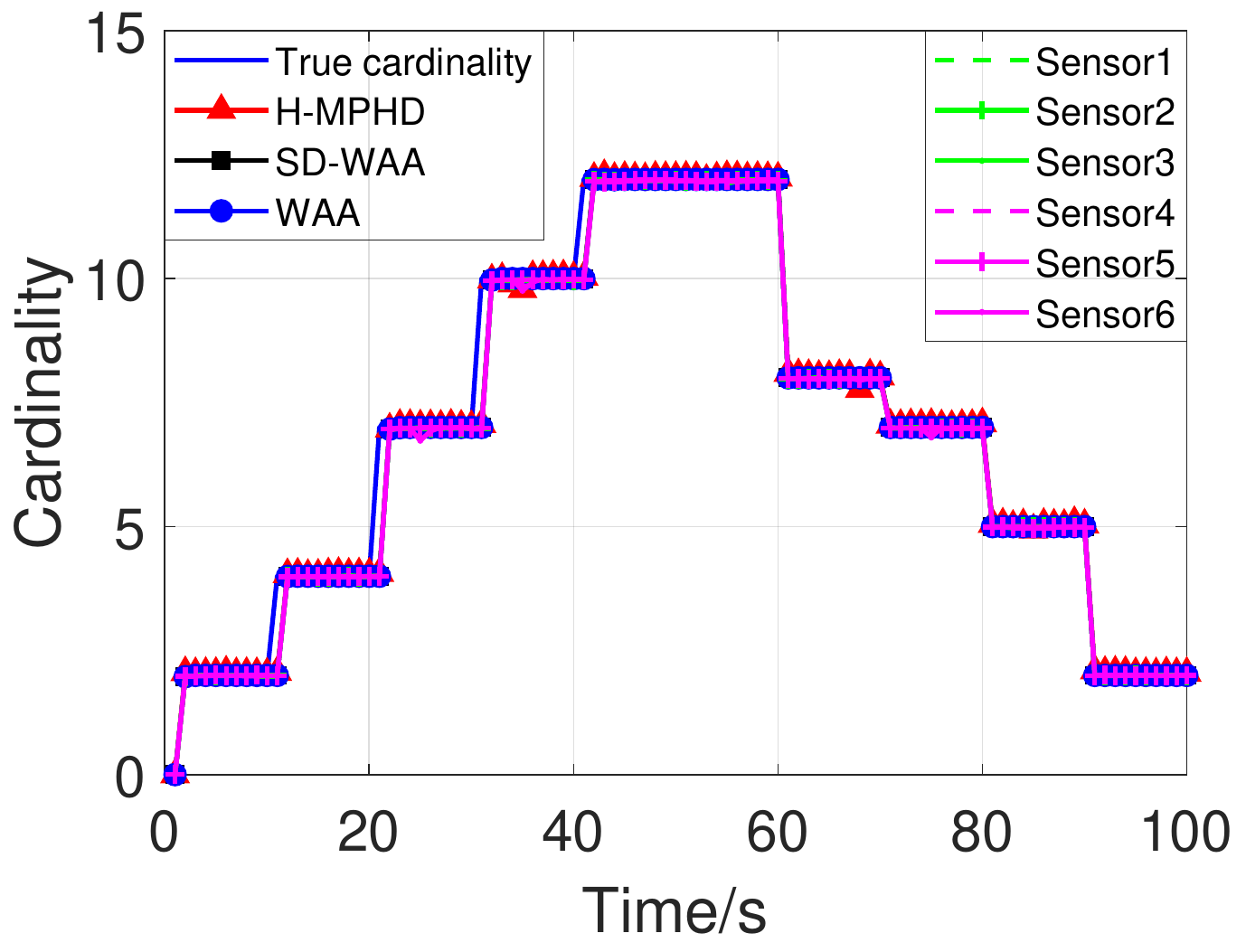}}
	\subfigure[]{\includegraphics[width=0.49\columnwidth,draft=false]{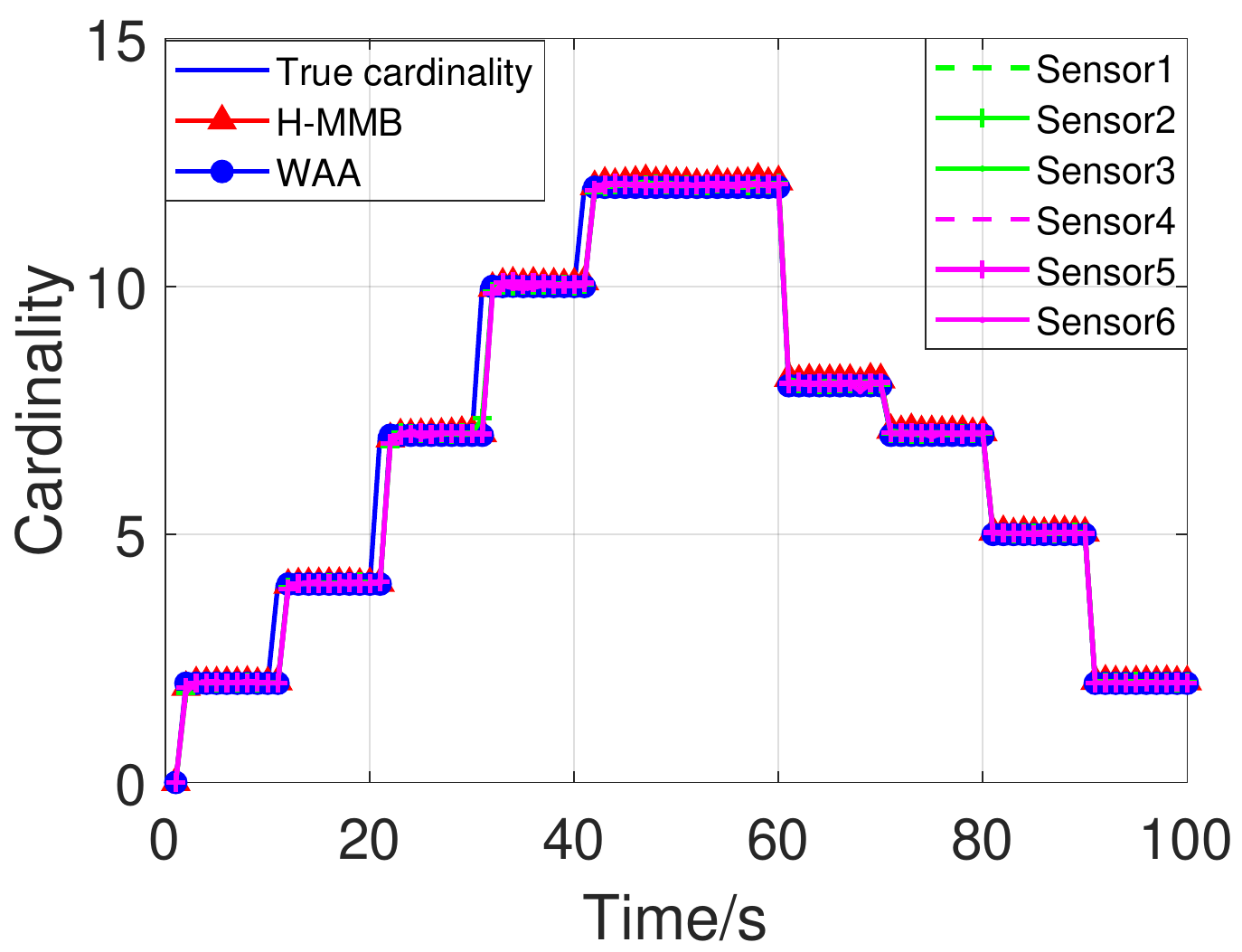}}
	\caption{{Case 2 : the cardinalities are plotted against time for all the single and multi-sensor PHD and MB filters, (a) the PHD filters, (b) the MB filters.}}
	\label{fig:ospa-phd-acc}
\end{figure}
\begin{figure}[t]
	\centering
	\subfigure[]{\includegraphics[width=0.49\columnwidth,draft=false]{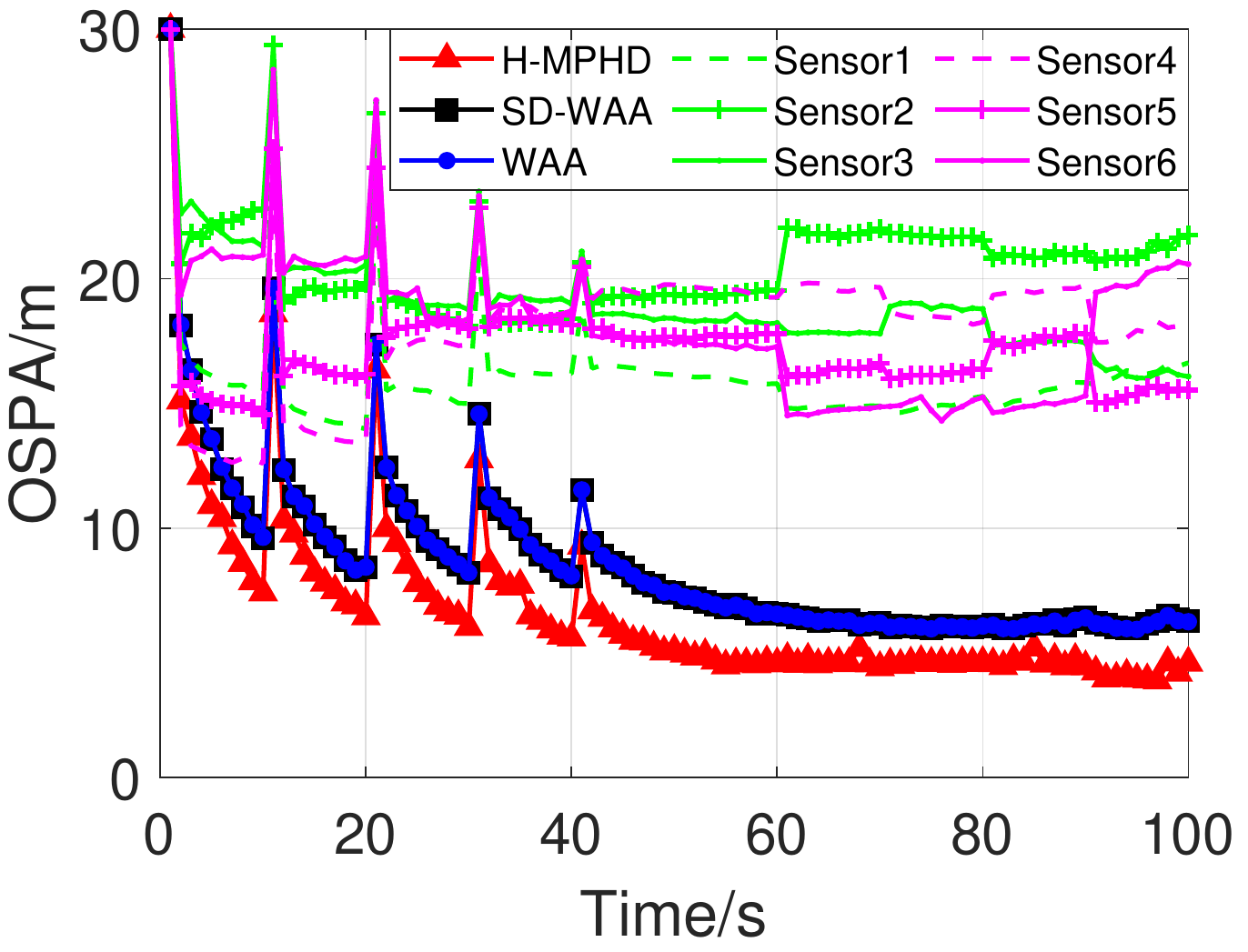}}
	\subfigure[]{\includegraphics[width=0.49\columnwidth,draft=false]{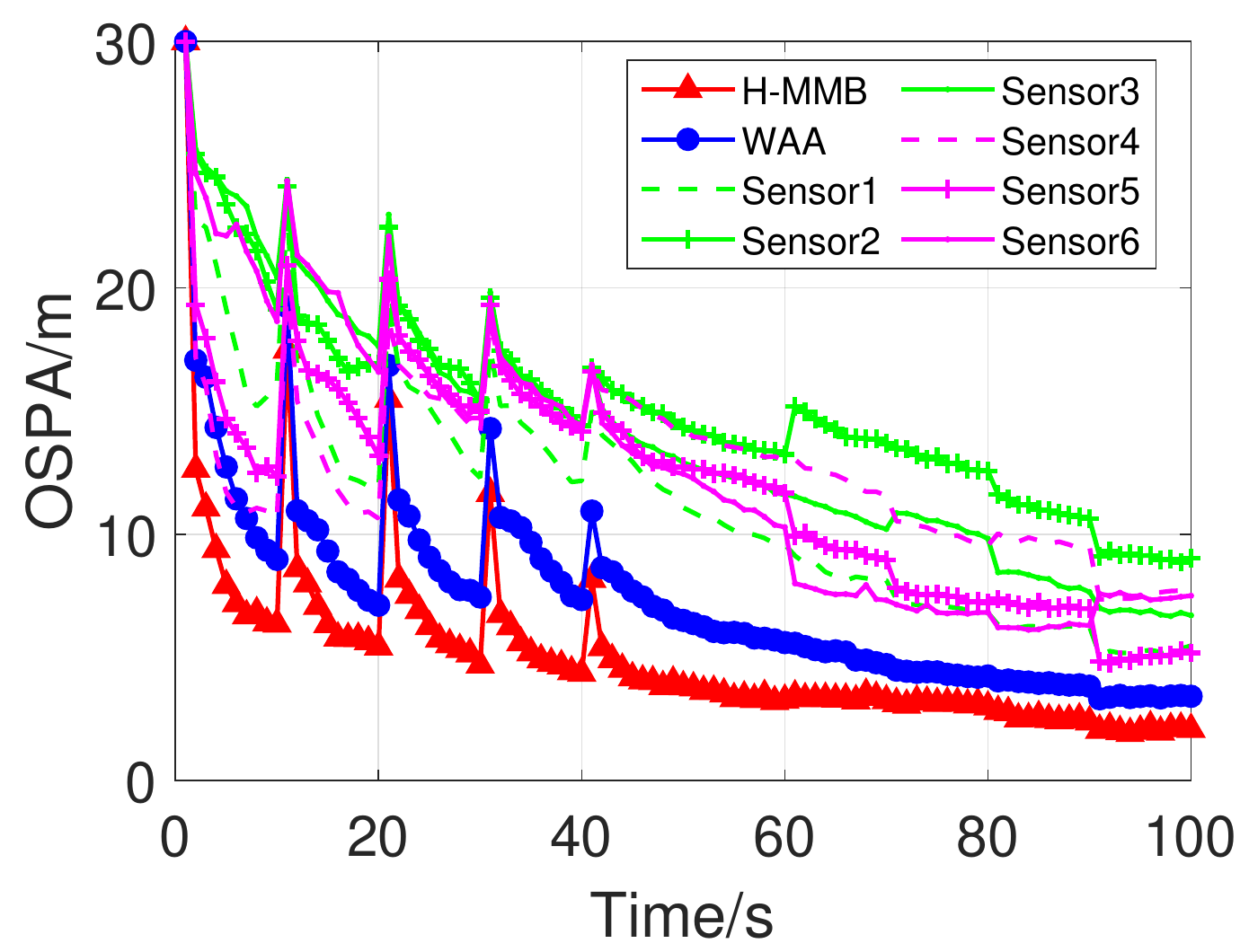}}
	\caption{{Case 2 : the OSPA errors are plotted against time for all the single and multi-sensor PHD and MB filters, (a) the PHD filters, (b) the MB filters.}}
	\label{fig:ospa-mb-acc}
\end{figure}
\begin{table}[t]
\begin{center}
			{\caption{The Averaged Execution Times {(\si{\second})} for One Time Step} \label{table3}
		\footnotesize
		\begin{tabular}{cccccc}
			\toprule
			\specialrule{0em}{1pt}{1.5pt}
			{Case} & {WAA-PHD} & {SD-WAA} &{H-MPHD} & {WAA-MB} & {H-MMB}  \\
			\specialrule{0em}{1pt}{1.5pt}
			\hline
			\specialrule{0em}{1pt}{1.5pt}
			{1} & $0.350$ & $0.353$ & $0.882$ & $0.439$ & $1.157$ \\
			\specialrule{0em}{1pt}{1.5pt}
			\specialrule{0em}{1pt}{1.5pt}
			{2} & $0.469$ & $0.478$ & $1.102$ & $0.513$ & $1.342$ \\
			\specialrule{0em}{1pt}{1.5pt}
			\bottomrule
		\end{tabular}}
	\end{center}
\end{table}
In Case 1, we show that the heterogeneous fusion strategy can better handle the sensors difference in terms of both cardinality and accuracy estimation abilities. In Case 2, we will study how heterogeneous fusion strategy performs when sensors only have different capabilities in estimation accuracy. Specifically, the sensor positions and ground truth of objects are the same as that in Case 1. The only difference is that the FoV of each sensor can completely cover the surveillance region, and the probability of detection is $p_D =0.98$ for all sensors over the whole surveillance area.

Firstly, the heterogeneous weights are plotted in Fig. \ref{fig:weight-acc}. It can be seen that the values of the weights are changing smoothly since only the AEUF $\Phi^a$ dominates the evaluation of the fusion weights. Note that the estimation accuracy of a sensor changes gradually since the value of $|R^x(z_m^\ast)|_\text{d}$ increases gradually as $d(s_i,x_m^\ast)$ increases due to the fixed angular measurement error. The cardinality and OSPA errors are plotted against time steps for each single sensor and all fusion methods in Figs. \ref{fig:ospa-phd-acc} and \ref{fig:ospa-mb-acc}. It can be seen that the proposed H-MPHD and H-MMB filters can still significantly outperform {the SD-WAA filter}, and the homogeneous fusion methods even the homogeneous ones have perfect cardinality estimation performance.

{Lastly, the averaged per time step execution times of all fusion algorithms are listed in Table \ref{table3} for both Case $1$ and Case $2$. The execution times were obtained using MATLAB (R2018b) on a laptop with a $3$ GHz Intel Core i3 processor and $8$ Gb memory. It can be seen that the proposed heterogeneous fusion algorithms have higher complexity because they involve extra processes such as the clustering of Bernoulli components and the assignment of individualized weights to Gaussian components. Besides, all fusion algorithms have longer execution times for Case $2$ since all local sensors have full FoVs that can cover the whole surveillance region.}

\section{Conclusion}
In this paper, we addressed the problem of the density based multi-sensor fusion using RFS type MODs. Firstly, we showed that the weighting mechanism of using a scalar coefficient in existing fusion methods can be oversimplified for practical scenarios. Then, we propose a novel heterogeneous fusion method, which adopts a \emph{local condition tailored} fusion tactic, to perform the information averaging among RFS MODs based on multiple sets of tailor-made fusion weights. To demonstrate the efficacy of the fusion strategy, we further derived the detailed fusion equations for both the PHD and MB filters. Simulation results confirmed the effectiveness and robustness of the proposed algorithms in challenging tracking scenarios.

{It should be noted that, compared to the existing homogeneous fusion methods, the additional information of environmental factors and sensor abilities are required to evaluate the heterogeneous fusion weights in the proposed algorithms. On the other hand, there is no principled method to incorporate sensor and environment information into the fusion process of the homogeneous fusion methods even if the information is available. Hence, the proposed heterogeneous method can be viewed as a principled extension of the homogeneous fusion methods by properly exploiting the sensor and environment information during the fusion process.}
{In a future study, we will consider the extension of the heterogeneous fusion method to the sensor networks with decentralized architecture~\cite{Uney2013,Battistelli2013,Yi2020TSP,Wang2017,Suqi2018,Sharma2019_TSP,Li2020_IEEESJ}.}

\appendices
\section{Proof of Proposition 1}\label{Prf:PoiPartition}
Using the property of conditional probability, the distribution ${\pi_m}$ of the $m$-th partitioned sub-set $X_m$ can be calculated as
\begin{equation}\label{eq:AP,divisor}
\begin{aligned}
{\pi_m}(X)&=\frac{{\pi}(X;X\subseteq{\mathbb{X}_m})}{\text{Pr}(X\subseteq{\mathbb{X}_m})}\\
&=\frac{{\pi}(X){I_{\mathbb{X}_m}(X)}}{\int_{{\mathbb{X}_m}} {\pi \left( X \right)\delta X}  },
\end{aligned}
\end{equation}
where the divisor in (\ref{eq:AP,divisor}) is a set integral on sub-space ${\mathbb{X}_m}$, and defined as,
\begin{equation}\label{eq:set integral}
\int \pi(X)\delta X = \frac{1}{n!} \int \pi(\{x_1,\ldots,x_n\}) dx_1,\ldots,dx_n.
\end{equation}

Recall that the MOD of an MPP RFS is given by~\cite{Mahler2007Book},
\begin{equation}\label{Poi distribution}
\pi \left( X \right) = {e^{ - \lambda }}\prod\limits_{x \in X}{\lambda}\cdot {f\left( x \right)}.
\end{equation}

Then, using (\ref{eq:set integral}) and (\ref{Poi distribution}), the set integral of an MPP RFS on sub-space ${\mathbb{X}_m}$ can be calculated as
\begin{equation}\label{eq:AP,int Poi}
\begin{aligned}
&\hspace{5mm}\int_{{\mathbb{X}_m}} {\pi \left( X \right)\delta X}\\
&= \sum\limits_{n = 0}^\infty  {\frac{1}{{n!}}\int_{(\mathbb{X}_m)^n} \pi \left( \{x_1,\ldots,x_n\} \right)  d{x_1},\ldots,d{x_n}}   \\
&= \sum\limits_{n = 0}^\infty  {\frac{1}{{n!}}\int_{(\mathbb{X}_m)^n} {{e^{ - \lambda }}{\lambda ^n}\prod\limits_{l = {1}} ^n {f\left( x_l \right)} } d{x_1},\ldots,d{x_n}}   \\
&= \sum\limits_{n = 0}^\infty  {\frac{1}{{n!}}{e^{ - \lambda }}{\lambda ^n}\prod\limits_{l = {1}} ^n {\int_{\mathbb{X}_m} {f\left( x_l \right)dx_l} } }   \\
&= {e^{ - \lambda }}\sum\limits_{n = 0}^\infty  {\frac{1}{{n!}}{\lambda ^n}{p_m}^n}   \\
&= {e^{ - \lambda }}{e^{\lambda  \cdot {p_m}}} . \\
\end{aligned}
\end{equation}

By substituting (\ref{Poi distribution}) and (\ref{eq:AP,int Poi}) into \eqref{eq:AP,divisor}, we have
\begin{align}
\nonumber
{\pi _m}\left( X \right) &= \frac{{{e^{ - \lambda }}{\lambda ^n}\prod\limits_{x \in {X}}} {f(x){I_{\mathbb{X}_m}(x)}} } {{e^{ - \lambda }}{e^{\lambda   {p_m}}} }  \\
&= {e^{ - \lambda   {p_m}}}  {\left( {\lambda   {p_m}} \right)^n} \cdot \prod\limits_{x \in {X}}\frac{f( x ) {I_{\mathbb{X}_m}(x)}}{{{p_m}}}   \\
\nonumber
&= {e^{ - {\lambda _m}}}{\lambda _m}^n\prod\limits_{x \in {X}} {f_m \left( x \right)}.
\end{align}
Thus, Proposition 1 is proved.

\section{Proof of Proposition 2}\label{Prf:Poi_Union}
The probability-generating functional (PGFl) of an MPP $X$ of the form (\ref{Poi distribution}) is given by~\cite{Mahler2007Book},
\begin{equation}\label{PHD pgfl}
{G_X}\left[ h \right] = {e^{\lambda f\left[ h \right] - \lambda }}
\end{equation}
where
\begin{equation}\label{eq:Poi_Union_fh}
f[h] = \int_{\mathbb{X}} h(x)f(x)dx
\end{equation}
and $h(x)$ is a nonnegative real-valued function of $x$ that has no units.

Since the random finite subsets $\{X_m\}_{m=1}^M$ are statistically independent, the PGFl of the RFS $X= \cup_{m=1}^M X_m$ can be expressed as the product of the PGFls of these subsets~\cite{Mahler2007Book},
\begin{equation}\label{eq:joint pgfl}
{G_X}\left[ h \right] = {G_{{X_1}}}\left[ h \right]\cdots{G_{{X_M}}}\left[ h \right].
\end{equation}

Using \eqref{PHD pgfl}, \eqref{eq:joint pgfl}, and the expression of $\bar{\pi}^{\text{\texttt{P}}}_m(X)$ (see eq. (18) in the main body of the manuscript),  we have
\begin{equation}\label{eq:merged Poi pgfl}
\begin{aligned}
{G_X}\left[ h \right] &= \prod\limits_{m = 1}^M {{e^{{{\lambda} _m}{{f}_m}\left[ h \right] - {{\lambda} _m}}}}   \\
&= {e^{ \left\{\sum\limits_{m =1}^M{{{\lambda}_m}{{f}_m}\left[ h \right]}-\sum\limits_{m = 1}^M {{{\lambda}_m}}\right\}}}\\
&= {e^{\left\{\breve \lambda \sum\limits_{m=1}^M {\frac{{{{\lambda}_m}}}{{\breve\lambda }}{{f}_m}\left[ h \right]}  - \breve \lambda \right\}}} \\
&= {e^{\breve \lambda{\breve{f}}\left[ h \right] - \breve \lambda }}
\end{aligned}
\end{equation}
Then, by comparing \eqref{eq:merged Poi pgfl} and \eqref{PHD pgfl}, the proof is concluded.

\section{The Expression of the Coordinate-converted Measurement Covariance Matrix}
Consider the state of an object is defined in Cartesian coordinate as $x=[p_{{\texttt{x}}},p_{\texttt{y}}]\in\mathbb{R}^2$. Let $z^x=[r^x,\theta^x]^{\top}$ denote the associated measurement of $x$ with $r^x$ and $\theta^x$ the range and angle measurements in polar polar coordinate, respectively. The standard deviations of the range and angle measurement noises are denoted as $\sigma_{r}$ and $\sigma_{\theta}$, respectively.
According to~\cite[(13a--13c)]{lerro1993tracking}, the coordinate-converted measurement covariance matrix has the following expression,
\begin{equation}\label{eq:transformed R}
R^{x}(z) = \left[ \begin{array}{{cc}}
R^{x}_{11} & R^{x}_{12} \\
R^{x}_{21} & R^{x}_{22}  \\
\end{array}
\right]
\end{equation}
where
\begin{align*}
R^{x}_{11} &= [r^{x}]^2e^{-2\sigma_{\theta}^2}\left[{\cos}^2\theta^{x}({\cosh}2\sigma_{\theta}^2-{\cosh}\sigma_{\theta}^2) \right. \\
&\qquad\qquad\left. +{\sin}^2\theta^{x}({\sinh}2\sigma_{\theta}^2-{\sinh}\sigma_{\theta}^2) \right] \\
& \quad+\sigma^2_{r}e^{-2\sigma_{\theta}^2} \left[{\cos}^2\theta^{x}(2{\cosh}2\sigma_{\theta}^2-{\cosh}\sigma_{\theta}^2) \right. \\
& \qquad\qquad \left. +{\sin}^2\theta^{x}(2{\sinh}2\sigma_{\theta}^2-{\sinh}\sigma^2_{\theta}) \right], \\
R^{x}_{22} &= [r^{x}]^2e^{-2\sigma_{\theta}^2}\left[\sin^2\theta^{x}({\cosh}2\sigma_{\theta}^2-{\cosh}\sigma_{\theta}^2) \right. \\
&\qquad\qquad\left. +\cos^2\theta^{x}({\sinh}2\sigma_{\theta}^2-{\sinh}\sigma_{\theta}^2) \right] \\
& \quad+\sigma^2_{r}e^{-2\sigma_{\theta}^2} \left[{\sin}^2\theta^{x}(2{\cosh}2\sigma_{\theta}^2-{\cosh}\sigma_{\theta}^2) \right. \\
& \qquad\qquad \left. +{\cos}^2\theta^{x}(2{\sinh}2\sigma_{\theta}^2-{\sinh}\sigma^2_{\theta}) \right], \\
R^{x}_{12} &= {\sin}\theta^{x}{\cos}\theta^{x}e^{-4\sigma_{\theta}^2}\left[\sigma_r^2+(1-e^{\sigma_{\theta}^2})\left([r^x]^2+\sigma_r^2\right)\right],\\
\end{align*}
and $R^{x}_{12} =R^{x}_{21}$.

\section{Proof of Corollary 2}\label{Prf:Cor_MB-PHD}
For a given MB MOD $\pi^{\text{\texttt{MB}}}_i$ parameterized by $\{(r_{i,b},f_{i,b}(x))\}_{b\in\mathbb{B}_i}$, its first order moment or PHD $\nu^{\text{\texttt{MB}}}_i(x)$ can be obtained via calculating the first order derivative of its PGFl \cite[(16.66)]{Mahler2007Book} as,
\begin{equation}\label{eq:PHD of MB MOD}
\begin{aligned}
\nu^{\text{\texttt{MB}}}_i(x) &= \left. \frac{\partial}{\partial x} G^{\text{\texttt{MB}}}_i [h] \right|_{h=1} \\
&= \left. \frac{\partial}{\partial x} \int h^X \breve\pi_i^{\text{\texttt{MB}}}(X) \delta X \right|_{h=1} \\
&= \sum\limits_{b\in\mathbb{B}_i} r_{i,b} f_{i,b}(x),
\end{aligned}
\end{equation}
where $G^{\text{\texttt{MB}}}_i [h]$ denotes the PGFl of the MB MOD $\pi^{\text{\texttt{MB}}}_i$.

Similarly, given a set of MB MODs and the associated homogeneous fusion weights $\{(\pi^{\text{\texttt{MB}}}_i,\omega_i)\}_{i\in\mathcal{N}}$, the PHD of the fused MOD $\breve{\pi}$ according to the direct AA fusion rule can be computed as,
\begin{equation}\label{eq:PHD of fused MB}
\begin{aligned}
\breve \nu^{\text{\texttt{MB}}}(x) &= \left. \frac{\partial}{\partial x} \int h^X \breve\pi(X) \delta X \right|_{h=1} \\
&= \left. \frac{\partial}{\partial x} \int h^X \sum\limits_{i\in {\mathcal{N}}} \omega_i \pi^{\text{\texttt{MB}}}_i(X) \delta X \right|_{h=1}.
\end{aligned}
\end{equation}
Next, using the linear property of set integral \cite[(11.102)]{Mahler2007Book} and (\ref{eq:PHD of MB MOD}), equation \eqref{eq:PHD of fused MB} can be further written as
\begin{equation}\label{eq:PHD of fused MB nom}
\begin{aligned}
\breve \nu^{\text{\texttt{MB}}}(x) &= \left. \sum\limits_{i\in\mathcal{N}} \omega_i \frac{\partial}{\partial x} \int h^X \breve\pi^{\text{\texttt{MB}}}_i (X) \delta X \right|_{h=1} \\
&= \sum\limits_{i\in\mathcal{N}} \sum\limits_{b\in\mathbb{B}_i} \omega_i r_{i,b} f_{i,b}(x).
\end{aligned}
\end{equation}

By substituting $\omega_{i,g}=\omega_i$ into the expression of  $\breve{\nu}^{\text{\texttt{MB}}}(x)$ (see eq. (45) in the main body of the manuscript), the PHD of the fused MOD by the H-MMB filter becomes,
\begin{equation}\label{eq:HMMB-PHD2}
\begin{aligned}
\breve{\nu}^{\text{\texttt{MB}}}(x)&=\sum\limits_{g=1}^{G_{\mathcal{C}}}\sum\limits_{i\in\mathcal{N}, {b\in \mathbb{M}_{i,g}}} \omega_{i,g} r_{i,b} f_{i,b}(x)\\
&= \sum\limits_{i\in\mathcal{N}} \sum\limits_{b\in\mathbb{B}_i} \omega_i r_{i,b} f_{i,b}(x).
\end{aligned}
\end{equation}
Thus, Corollary 2 is proved by comparing \eqref{eq:PHD of fused MB nom} and \eqref{eq:HMMB-PHD2}.

\bibliographystyle{IEEEtran}
\bibliography{HWAA}

\end{document}